\documentstyle[12pt,aaspp4]{article}

\def\eps@scaling{.95} \def\epsscale#1{\gdef\eps@scaling{#1}}

\def\plotone#1{\centering \leavevmode
  \epsfxsize=\eps@scaling\columnwidth \epsfbox{#1}}

\def\plotfiddle#1#2#3#4#5#6#7{\centering \leavevmode \vbox
  to#2{\rule{0pt}{#2}} \includegraphics{#1}}

\def\spose#1{\hbox to 0pt{#1\hss}} \def\simlt{\mathrel{\spose{\lower
      3pt\hbox{$\mathchar"218$}} \raise 2.0pt\hbox{$\mathchar"13C$}}}
\def\simgt{\mathrel{\spose{\lower 3pt\hbox{$\mathchar"218$}} \raise
    2.0pt\hbox{$\mathchar"13E$}}} \def\lsim{\rlap{$<$}{\lower
    1.0ex\hbox{$\sim$}}} \def\gsim{\rlap{$>$}{\lower
    1.0ex\hbox{$\sim$}}}   \def\kms{\mbox{{\rm
      km~s}$^{-1}$}}

\newcommand{\cms}{\mbox{cm$^{-2}$}}
\newcommand{\Lya}{\mbox{Ly$\alpha$}}
\newcommand{\lya}{\mbox{Ly$\alpha$}}
\newcommand{\Lyb}{\mbox{Ly$\beta$}}
 
\newcommand{\NHI}{\mbox{$N_{\mathrm HI}$}} 

\newcommand{\hi}{\mbox{H {\sc i}}} 
 
\newcommand{\heii}{\mbox{He {\sc ii}}}

\newcommand{\ovi}{\mbox{O {\sc vi}}}

\lefthead{Heap, Williger, Smette et al.}
\righthead{The \ion{He}{2} Gunn-Peterson Effect}

\def \m.  {\rlap{$.$}^{\rm m}} \def \s.  {\rlap{$.$}^{\rm s}} \def
\am.  {\rlap{$.$}'} \def \as.  {\rlap{$.$}''}

\def \deg.  {\rlap{$.$}^\circ}

\newcommand{\fcaption}[1]{ \refstepcounter{figure} \setbox\@tempboxa =
  \hbox{\tenrm Fig.~\thefigure. #1} \ifdim \wd\@tempboxa > 6in
  {\begin{center} \parbox{6in}{\tenrm\baselineskip=12pt
        Fig.~\thefigure. #1 }
            \end{center}}
          \else {\begin{center} {\tenrm Fig.~\thefigure. #1}
              \end{center}}
            \fi}

          \newcommand{\tcaption}[1]{ \refstepcounter{table}
            \setbox\@tempboxa = \hbox{\tenrm Table~\thetable. #1}
            \ifdim \wd\@tempboxa > 6in {\begin{center}
                \parbox{6in}{\tenrm\baselineskip=12pt Table~\thetable.
                  #1 }
            \end{center}}
          \else {\begin{center} {\tenrm Table~\thetable. #1}
              \end{center}}
            \fi}
            
\def\NHEII{\mbox{$N_{\mathrm He II}$}}

\def\NHI{\mbox{$N_{\mathrm H I}$}}

{\setcounter{table}{0}

\slugcomment{accepted for publication in ApJ}

\begin{document}
\bibliographystyle{unsrt}
\pagenumbering{arabic}


\title{STIS Observations of \ion{He}{2} Gunn-Peterson Absorption \\
        Toward  Q~0302--003 $^\dagger$}

\vspace{1cm}

\author{Sara R. Heap$^{1,5}$, Gerard M. Williger$^{1,2,6}$, 
Alain Smette$^{1,2,7}$, Ivan Hubeny$^{1,2,8}$, \\
Meena S. Sahu$^{1,2,9}$, 
Edward B. Jenkins$^{3,10}$, Todd M. Tripp$^{3,11}$,
Jonathan N. Winkler$^{4,12}$}

\medskip
\noindent
$^{\phantom{1}1}$ Laboratory for Astronomy \& Solar Physics, Code 681, NASA/GSFC,
       Greenbelt MD~20771, USA\\
$^{\phantom{1}2}$ National Optical Astronomy Observatories, Tucson, AZ 85726, USA\\
$^{\phantom{1}3}$ Princeton University Observatory, Princeton, NJ 08544, USA\\
$^{\phantom{1}4}$ Queen's College, Oxford OX1 4AW, England\\
$^{\phantom{1}5}$ heap@srh.gsfc.nasa.gov\\
$^{\phantom{1}6}$ williger@tejut.gsfc.nasa.gov\\
$^{\phantom{1}7}$ asmette@band3.gsfc.nasa.gov\\
$^{\phantom{1}8}$ hubeny@tlusty.gsfc.nasa.gov\\
$^{\phantom{1}9}$ msahu@panke.gsfc.nasa.gov\\
$^{10}$ ebj@astro.princeton.edu\\
$^{11}$ tripp@astro.princeton.edu\\
$^{12}$ jonathan.winkler@queens.oxford.ac.uk\\

\medskip

\noindent $\dagger $Based on observations with the NASA/ESA Hubble Space
Telescope, obtained at the Space Telescope Science Institute, which is
operated by the Association of Universities for Research in Astronomy, Inc.,
under NASA contract NAS5-26555.

\begin{abstract}

The ultraviolet spectrum (1145--1720~\AA) of the distant quasar
Q~0302--003 ($z=3.286$) was observed at 1.8~\AA\, resolution with the
Space Telescope Imaging Spectrograph aboard the Hubble Space
Telescope. A total integration time of 23,280 s was obtained.  The
spectrum clearly delineates the Gunn-Peterson \ion{He}{2} absorption
trough, produced by \ion{He}{2}~\Lya\, along the line of sight over
the redshift range $z=2.78 - 3.28$.  Its interpretation was
facilitated by modeling based on Keck HIRES spectra of the \ion{H}{1}
\Lya\, forest (provided by A. Songaila and by M. Rauch and W. Sargent).  
We find that near the quasar, \ion{He}{2} \Lya\, absorption is
produced by discrete clouds, with no significant diffuse gas; this is
attributed to a \ion{He}{2} ``proximity effect'' in which the quasar
fully ionizes He in the diffuse intergalactic medium, but not the He
in denser clouds.  By two different methods we calculate that the
average \ion{He}{2} \Lya\, opacity at $z \approx 3.15$ is $\tau \geq
4.8$.  In the Dobrzycki-Bechtold void in the \ion{H}{1} \Lya\, forest
near $z=3.18$, the average \ion{He}{2} opacity
$\tau=4.47^{+0.48}_{-0.33}$.  
Such large opacities require the presence of a diffuse gas component
as well as a soft UV background spectrum, whose softness parameter,
defined as the ratio of the photo-ionization rate in \ion{H}{1} over
the one in \ion{He}{2} $S \equiv \Gamma^{\mathrm J}_{\mathrm
  HI}/\Gamma^{\mathrm J}_{\mathrm HeII}\simeq 800$, indicating a
significant stellar contribution.  At $z=3.05$, there is a distinct
region of high \ion{He}{2} \Lya\, transmission which most likely
arises in a region where helium is doubly ionized by a discrete local
source, quite possibly an AGN.
At redshifts $z<2.87$, the \ion{He}{2} \Lya\, opacity detected by
STIS, $\tau=1.88$, is significantly lower than at $z>3$.  Such a
reduction in opacity is consistent with Songaila's (1998) report that
the hardness of the UV background spectrum increases rapidly from $z
=3$ to $z = 2.9$.

\end{abstract}


\newpage

\section{INTRODUCTION}
\label{sec:introduction}

Once-ionized helium is the most abundant absorbing ion in the
intergalactic medium (IGM).  It outnumbers \ion{H}{1} by a factor
equal to the ratio of the ion number densities $n_{\mathrm
  HeII}/n_{\mathrm HI} \ga 100$, and therefore serves as an  ideal
tracer of the IGM in the early universe ($z>2$). The presence of
\ion{He}{2} is signaled by absorption of \ion{He}{2} \Lya\, $\lambda
304$~\AA\, redshifted to the far-ultraviolet (far-UV) where it can be
observed by space observatories such as HST. If \ion{He}{2} ions
reside in clumps or clouds positioned along the line of sight to a
quasar, they produce discrete absorption lines similar to the
\ion{H}{1} Ly$\alpha$ forest.  But if instead they are diffused
throughout the IGM, they will smoothly depress the flux level of a
quasar shortward of its \ion{He}{2} Ly~$\alpha$ emission line. This is
known as the Gunn-Peterson effect (\cite{Gunn65}; \cite{Scheuer65}).
Although the Gunn-Peterson opacity was originally formulated for
\ion{H}{1}, there is an analogous formula for \ion{He}{2}:
\begin{equation}
\tau _{\mathrm{HeII}}(z)
=
\frac{c}{H_{\mathrm o}} ~ 
\frac{n_{\mathrm{HeII}}(z) ~ \sigma _{\mathrm{HeII}}}
{(1+z) \sqrt{1+\Omega _{\mathrm o} z}} ~,
\end{equation}
where $\sigma_{\mathrm HeII} \equiv (\pi e^{2}/m_{e}c^2) f \lambda$  is
the resonant scattering \Lya\,  cross-section, and $f$ and $\lambda$ are
the oscillator strength and wavelength of the \ion{He}{2} \lya\
line, respectively.  The cross-section for \ion{H}{1} is a factor of 4
larger than for \ion{He}{2}. Hence, the ratio of opacities, 
$R = \tau_{\mathrm{HeII}}(z)/\tau _{\mathrm{HI}}(z) = \eta /4 $,
where $\eta \equiv N_{\mathrm  HeII}/N_{\mathrm HI}$, and
$N_{\mathrm  HI}$ and $N_{\mathrm HeII}$ are the column densities
of \ion{H}{1} and \ion{He}{2}, respectively. In evaluating
the Gunn-Peterson optical depth and throughout this paper, we assume
$H_{\mathrm o} = 65~{\mathrm \kms}~ {\mathrm Mpc}^{-1}$, 
$\Omega_{\mathrm o}=0.2$ and $\Lambda=0$.

\ion{He}{2} Gunn-Peterson absorption has been explored in a variety of
numerical simulations (e.g. \cite{Zheng95}; \cite{Miralda96};
\cite{Croft97}; \cite{Zhang98}; \cite{Miralda99}, hereafter MHR). 
It has also been
observed in four lines of sight.  \ion{He}{2} Gunn-Peterson absorption
was first detected in the spectrum of the $z=3.286$ quasar,
Q~0302--003, obtained with the Faint Object Camera (Jakobsen et al.
1994).  However, because of the low spectral resolution and
signal-to-noise ratio, the FOC data could not distinguish
whether the absorption is produced by a uniformly distributed medium
or by the \ion{He}{2} counterpart of a forest of \Lya\, lines
(cf. Songaila et al. 1995).  Subsequent spectra covering the
wavelength range, 1140--1530~\AA\,  taken by the Goddard High Resolution
Spectrograph (Hogan et al. 1997) confirmed the strongly depressed flux
level shortward of \ion{He}{2} Ly$\alpha $ in the QSO rest frame. It
also showed a ledge in the flux level between 1280 -- 1295~\AA ,
attributed to the proximity effect (\cite{Bajtlik88}; \cite{Zheng95};
\cite{Giroux95}), an effect of increased ionization of gas in the
vicinity of the QSO.  In the region from 1240 to 1280~\AA , the GHRS 
spectrum was consistent with a null flux -- however one with
unacceptably large uncertainties (cf. Heap 1997).

Measurements of the \ion{He}{2} Gunn-Peterson effect provide a test of
cosmological models. Cosmological simulations can make concrete
predictions of the epoch of IGM re-ionization from singly to doubly
ionized helium (e.g. MHR), the timescale for the
ionization change, and the characteristics of 
transition regions which might produce patchy \ion{He}{2} absorption
(\cite{Reimers97}).  Combining observations and simulations would
constrain the density and ionization state of the gas, and the diffuse far-UV
background radiation at $z\sim 3$.

In this paper, we present new observations of Q~0302--003 made with
the Space Telescope Imaging Spectrograph (STIS). STIS brings several
improvements over GHRS for faint, point-source spectroscopy: first, a
lower instrumental background resulting in higher S/N ratios; second,
an imaging format that allows direct measurement of the (sky +
instrumental) background; and finally, a wider spectral range covered
in a single exposure, presenting an opportunity to study the
properties of the IGM over a broader range in redshift. The STIS
observations presented here permit the strictest constraint to date
for \ion{He}{2} Gunn-Peterson absorption, over a wide spectral interval,
and thus offer a significant advance in our understanding of the
physical processes at $z\sim 3$. In \S\ref{sec:observations} and
\S\ref{sec:observationalresults}, we describe the new observations. We
then analyze in detail three important aspects of Gunn-Peterson
absorption: the opacity of the absorption trough (\S\ref{sec:heiigptrough}), 
the ``proximity effect'' (\S\ref{sec:modeling}), and 
the appearance of opacity gaps in an otherwise solid wall of absorption 
(\S\ref{sec:voidz305}).  
In \S\ref{sec:discussion}, we summarize the implications of our findings.

\section{OBSERVATIONS AND REDUCTIONS}
\label{sec:observations}

Our analysis of \ion{He}{2} Gunn-Peterson absorption along the line of
sight to Q~0302--003 relies on two sets of observations: a far-UV
spectrum of the QSO taken with STIS at low resolution, and
high-resolution Keck spectra obtained by Songaila and by Rauch and
Sargent, which were kindly made available to us. In this section, we describe
the two datasets and our techniques of data reduction.

\subsection{STIS UV Spectra}
\label{sec:stis_uv_spectra}
We obtained STIS observations of Q~0302--003 in December 1997 during
the course of two ``visits,'' each five orbits long (Program 7575).
See { Woodgate et al. (1998) and Kimble et al. (1998)} for details of
the instrument and its performance. The observation used the G140L
grating which produces a spectrum covering the wavelength interval,
1145--1720~\AA , at a nominal two-pixel resolution of 1.2~\AA .  The
sensitivity decreases strongly with wavelength below its peak at
1300~\AA: it is only 10\% of its peak value at 1150~\AA , less than
4\% at 1145~\AA , and zero at at 1140~\AA .  The total exposure time
(10 orbits) was 23,280 s.

Because of the faintness of the QSO, we selected a rather wide
entrance slit (0\farcs2) in order to maximize throughput. As a
consequence, the resolution is degraded to about 3 pixels, or 1.8~\AA.
A wide slit also has the disadvantage of transmitting rather strong
geo-coronal emission of \ion{H}{1} \Lya\,  and a triplet of \ion{O}{1}
centered at $\lambda 1302$ (cf.  Figure 1). This can be a serious
problem because \Lya\,  lies in the Gunn-Peterson trough, and \ion{O}{1}
$\lambda 1302$ lies at its edge and might therefore interfere with the
assessment of the ``proximity effect''. To counteract this problem, we
obtained the spectra in the ``TIME-TAG'' mode with the intention of
using only those data recorded during the earth-shadow time.  We
found, however, that contamination by geo-coronal emission can be
fully accounted for in the reduction. We therefore assembled all of
the TIME-TAG data from each orbit into a conventional spectrogram.

\begin{figure}[htp]
  \begin{center}
    \plotfiddle{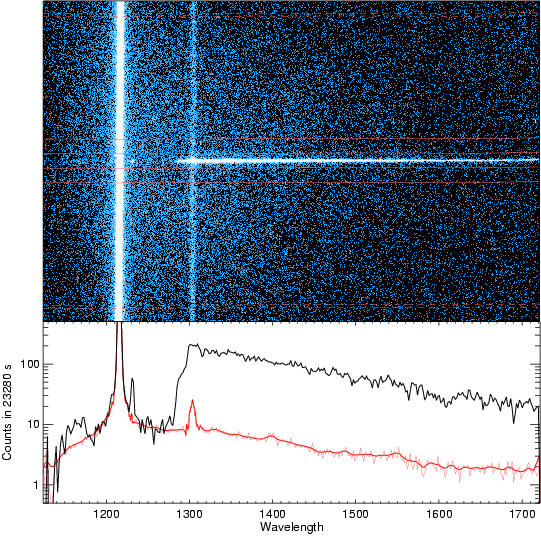}{10cm}{0}{70}{60}{-220}{-20}
    \caption{STIS observations of Q~0302--003. The top
  panel shows a major portion of the co-added spectrogram (lines
  310-970 on the detector format).  The two bright emission lines are
  geo-coronal Lyman $\alpha $ (left) and OI $\lambda $1302 (right).
  The spectrum of the QSO has a strong continuum flux at $\lambda
  \gtrsim 1280$~\AA , while the Gunn-Peterson absorption trough sets
  in at shorter wavelengths.  Within this trough, there is a major
  flux finger at $\lambda $1231, and several minor regions of flux are
  visible shortward of Lyman $\alpha $. The dashed red lines delimit
  the regions used to determine the background in the STScI reduction
  (not used for this paper), while the solid red lines apply to the
  region we adopted.  The bottom panel shows the extracted spectrum,
  including the gross spectrum (QSO + background) shown by the black
  line and the average of the background above and below the quasar
  (squiggly red line). In the reduction, the background was smoothed
  (smooth orange line).  Since the STIS MAMA1 detector is a
  photon-counting imaging detector, the errors in the total counts are
  simply the square root of the counts.\label{fig:srhfig1}}
  \end{center}
\end{figure}

The observations were reduced at the Goddard Space Flight Center with
the STIS Investigation Definition Team (IDT) version of CALSTIS
(Lindler 1998), which allowed us to make a customized treatment of the
background. Such flexibility is needed in order to ensure accurate
fluxes and opacities in the Gunn-Peterson trough. As shown in {
  Figure~\ref{fig:srhfig1}}, the background in a STIS G140L
spectrogram is highly non-uniform.  The strong geo-coronal emission at
Ly$\alpha $ and \ion{O}{1}~$\lambda$1302 affects a region wider
than the projected entrance slit because of grating scatter. An
additional source of background is the bright, diffuse background at
the short-wavelength region of the image (the MAMA1 ``hotspot''; cf.
Landsman 1998). The QSO spectrum intersects this hotspot roughly
through the middle.  Because of the non-uniformity of the background,
we sampled the background in two zones on either side of the QSO
spectrum as closely as possible to the QSO (30-pixel offset) using an
extraction slit height 30 pixels high (red lines in
Figure~\ref{fig:srhfig1}). In order to avoid smearing out the
geo-coronal emission lines, we smoothed the background (15-point
running mean, executed twice) only in the regions away from the
emission lines. In contrast, the STScI data processing system (RSDP)
samples the background of STIS G140L spectra within bands located 
$\pm 300$ pixels away from the QSO spectrum (inside each pair of dashed red 
lines in Figure~\ref{fig:srhfig1}) where the background is only  
half that of  our measured background. 
The resulting spectrum shows a spurious
residual flux in the Gunn-Peterson absorption trough.

\begin{figure}[htp]
  \begin{center}
    \plotfiddle{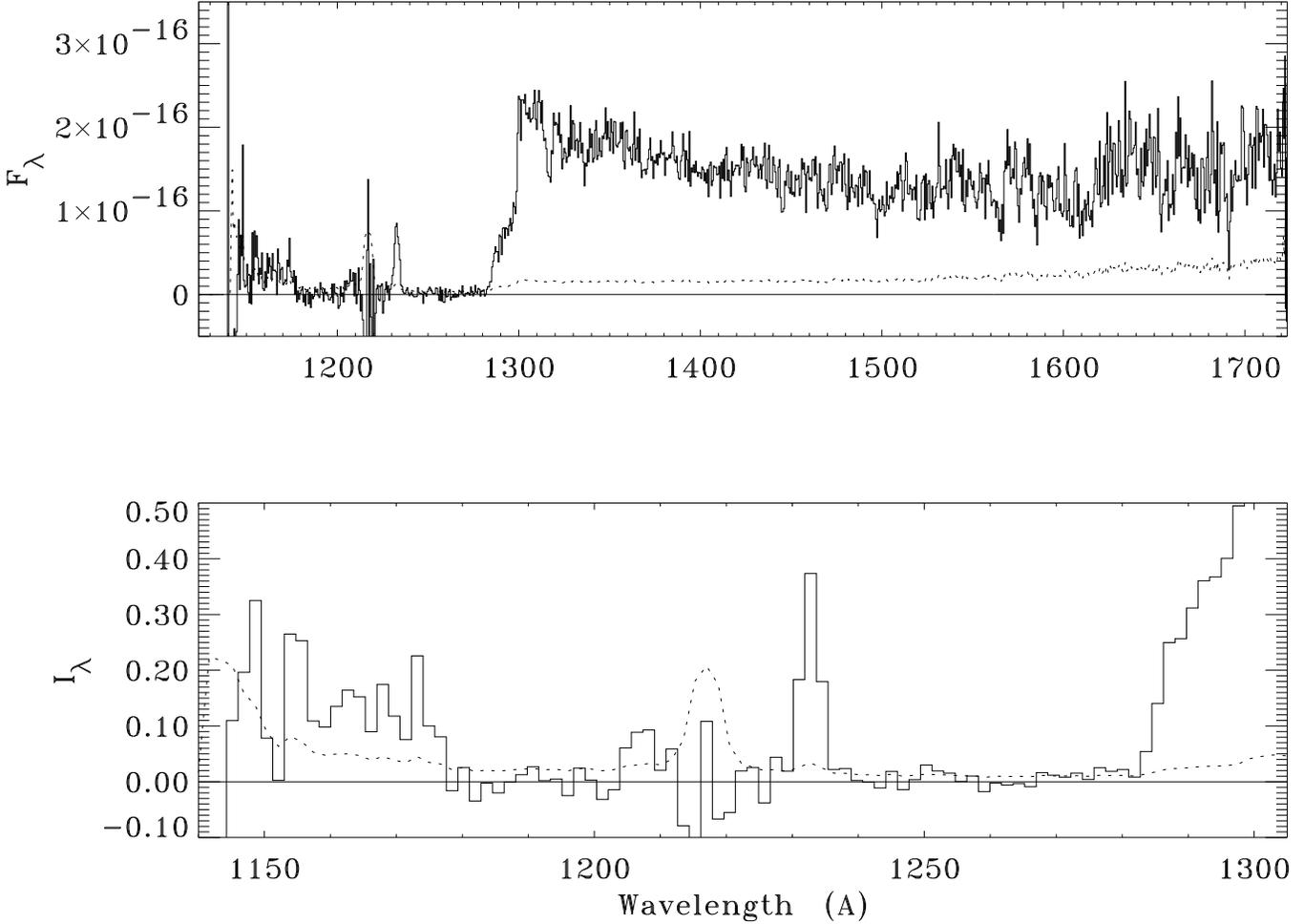}{12cm}{0}{110}{110}{-360}{-400}
    \caption{The STIS spectrum of Q~0302--003 after applying
  a background correction. Top: the absolute flux distribution in erg
  s$^{-1}$ cm$^{-2}$ \AA$^{-1}$ (solid line) showing \ion{He}{2}
  absorption shortward of 1300~\AA , overlaid with the 1 $\sigma$
  uncertainty (dashed line).  Bottom: a magnified portion of the
  spectrum in the Gunn-Peterson absorption trough.  In this panel, the
  flux was normalized to unity at $F_{\mathrm c} = 2.1\times 10^{-16}$
  erg s$^{-1}$ cm$^{-2}$ \AA$^{-1}$, and binned by 3 pixels in order
  to show the average flux in a resolution element.
\label{fig:srhfig2}}
  \end{center}
\end{figure}

Figure~\ref{fig:srhfig2} shows our reduced spectrum overplotted with
an error spectrum that assumes $\sqrt{N}$ statistics. 
Despite the low S/N ratio at short wavelengths, the residual flux
below 1175~\AA\, is real, since we see repeated appearances of signals
at about the $5\sigma$ level. There is also a feature just shortward of
geocoronal \Lya\,  detected at the $5\sigma$ level.  The wavelength
uncertainty is much harder to estimate. We made an on-board
calibration between successive exposures, which ensures
that the dispersion is precise.  However, there is still the
possibility of a shift in the spectral direction between the QSO
spectrum and the spectrum of the calibration source. Such a shift
might be induced by imperfect centering of the QSO within the 0\farcs2
slit (8 pixels wide) or by thermally-induced drifts between the QSO
and calibration exposures. Indeed, we find a 3-pixel shift in the zeropoint of
the wavelength scale between the first exposure in a visit and the other
four. Because of these uncertainties, we allow for a $\pm$1 pixel
(0\farcs025) offset in the observed wavelength scale.  
Consistency with the wavelength scale of the \ion{H}{1} Lyman forest
suggests that a 1.06-\AA\, shift (1.8 pixels) is in order.

The spectral resolution is also difficult to estimate because there
are no sharp, well defined lines in the co-added spectrum. We
therefore used the cross-dispersion profile of the co-added QSO
spectrum as a proxy for the line spread function.  The
cross-dispersion profile grows broader toward shorter wavelengths. We
estimate that in the Gunn-Peterson absorption trough, the FWHM 
is 3.1 pixels; thus the corresponding spectral resolution is
1.80~\AA.

\subsection{Keck HIRES Spectra}
\label{sec:keckhiresspectra}

We used spectra of Q~0302--003 taken with the Keck HIRES spectrograph
({Vogt et al. 1994}) provided by A. Songaila and by M. Rauch and W.
Sargent to compare \ion{H}{1} \Lya\,  absorption with the \ion{He}{2}
\Lya\,  absorption in the STIS spectrum.  {Table~\ref{tab:groundspec}}
summarizes the data.  Successive columns give the observed wavelength
range of the echellogram in \AA , the equivalent wavelength range
applicable to the STIS spectrum (i.e. the observed wavelength divided
by 4 for the \ion{He}{2} /\ion{H}{1} wavelength ratio), the observed resolution in
\kms, the maximum signal-to-noise ratio, and observer/source of
the data.  Hu et al. (1995) and Rauch et al. (1999) provide reduction
details for the Songaila and Rauch datasets respectively.  We
processed the Rauch data and estimated the 1$\sigma $ uncertainties
from the detector noise and photon counting statistics; for the
Songaila data we calculated the 1$\sigma$ uncertainties from the apparent {\it rms}
noise in the spectrum itself.  In our analysis, we relied on the Rauch spectrum at
$\lambda \simgt 4600$ \AA\,  where it has higher S/N ratio; we used the
Songaila spectrum at shorter wavelengths for some metal-line
identifications and to improve our \ion{H}{1} \lya\ profile
fits using higher members of \ion{H}{1} Lyman series.  We detected a 2--3\%
residual flux in the bottoms of saturated absorption lines in the
Songaila data, but found that it makes no significant difference
to our analysis.

\begin{center}

\begin{table}[ht]
\caption{Keck HIRES Spectra of the \ion{H}{1} Lyman Forest Toward Q~0302--003}
\label{tab:groundspec} 
\begin{tabular}{crccl}
\hline\hline
$\lambda_{\mathrm vac}$ (\AA )  & \multicolumn{1}{c}{$\lambda/4$} & $R$ (km/s) & 
$(S/N)_{\mathrm{max}}$ & Observer \\  \hline
3650--5983 &   912-1496 & 8.3 &\phantom{1}70  & Songaila \\  
4220--6656 &  1050-1664 & 6.9 & 130  & Rauch \& Sargent \\  \hline
\end{tabular}
\end{table}
\end{center}

Absorption lines of \ion{H}{1}, \ion{C}{4} and other ions detected in
the Keck spectra were fitted with Voigt profiles with the program,
VPFIT (Webb 1987). We base our redshifts on wavelengths corrected to
the vacuum heliocentric frame.  Higher members of the \ion{H}{1} Lyman
series were used wherever possible to constrain the profile fits.  We
estimate the completeness limit at $\log N_{\mathrm HI}\sim 13.0$ (\cms )
based on the point at which the \ion{H}{1} column density distribution
turns over.  Our results for redshift $z$, Doppler parameter $b_{\mathrm
  HI}$, and \ion{H}{1} column density, $N_{\mathrm HI}$, largely agree
with those published by Hu et al. (1995).  The few differences that we
find can be attributed to $(i)$ the higher resolution of the Rauch
dataset, $(ii)$ newly identified metal lines, or $(iii)$ an accounting
for the ink spot in the middle of the HIRES CCD.  The absorption-line
parameters for the \ion{H}{1} systems will be used in \S\ref{sec:modeling} to model
the \ion{He}{2} absorption.
A detailed line list will be presented elsewhere
(\cite{Rauch99}).

\section{OBSERVATIONAL RESULTS}
\label{sec:observationalresults}

In this section, we present the STIS spectrum of Q~0302--003, which
covers the wavelength interval, 267--402~\AA\, in the rest frame of
the QSO. We make this
presentation in the form of two comparisons. In \S3.1, we compare the
\ion{He}{2} \Lya\,  absorption trough ($\lambda_{\mathrm rest} < 304$~\AA )
with the corresponding \ion{H}{1} \Lya\,  forest. In \S3.2, we compare
the \ion{He}{2} \Lya\,  absorption in the spectrum of Q~0302--003 with
other lines of sight.

\subsection{Comparison of the \ion{He}{2} and \ion{H}{1} \Lya\,  Spectra}

\begin{figure}[htp]
  \begin{center}
    \plotfiddle{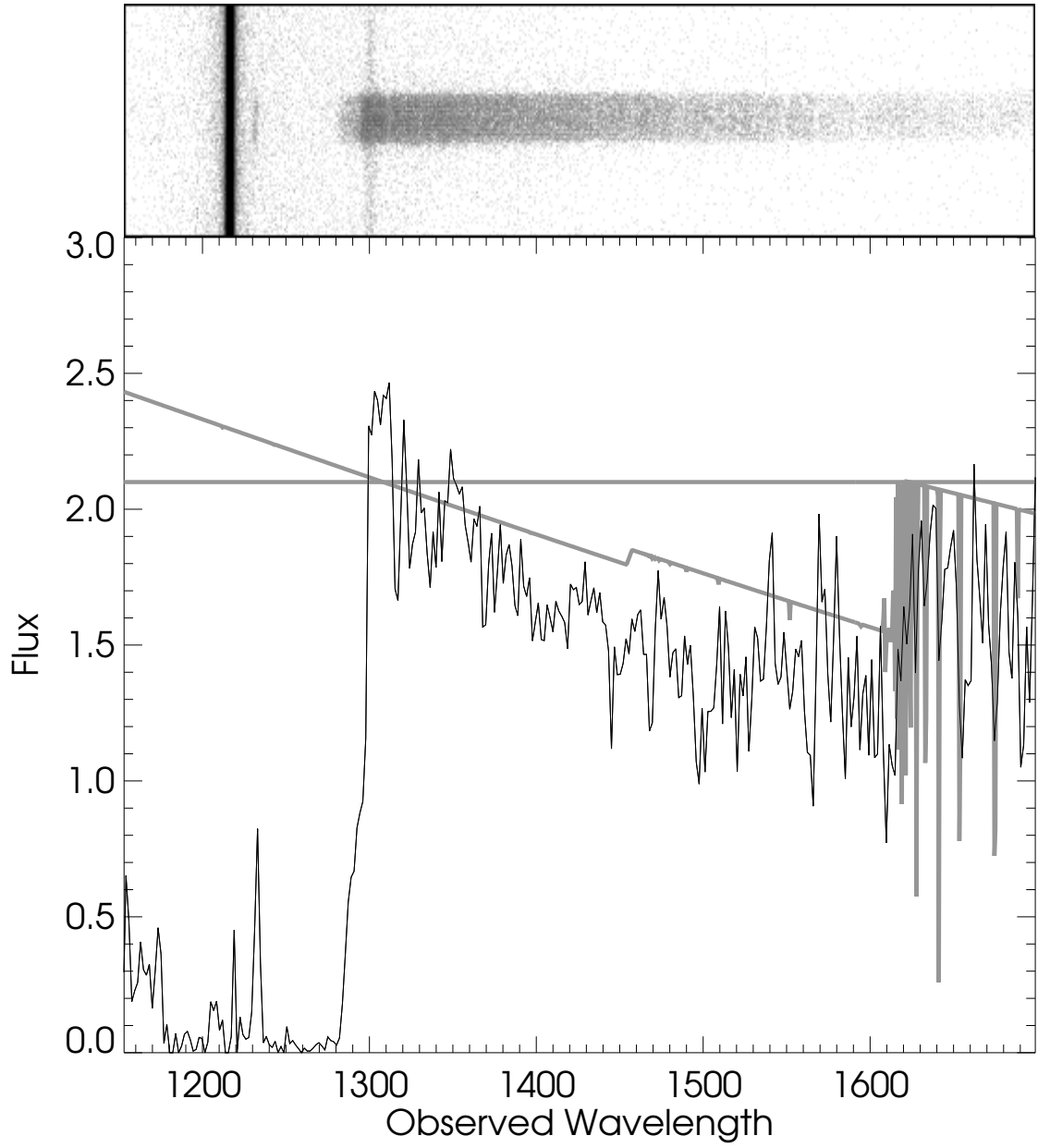}{16cm}{0}{120}{120}{-310}{-90}
    \caption{Far-UV flux distribution of Q~0302--003. Top: a 2-D display of
the 10 spectra stacked vertically to show the reality of spectral features.
Bottom: the observed flux distribution in units of erg s$^{-1}$ cm$^{-2}$ 
\AA$^{-1}$. Two continuum estimates are shown as gray lines: one allowing
for a Lyman limit system at $z=0.773$, and the other set to a constant value.
The latter was used to normalize the flux in the Gunn-Peterson trough.
\label{fig:srhfig3}}
  \end{center}
\end{figure}

In order to measure the strength of \ion{He}{2} \Lya\,  absorption, it
is necessary first to define the continuum flux distribution of the
QSO in this spectral region.  Of course, the continuum is not a
directly observable quantity, since it has been blocked by \ion{He}{2}
\Lya\ absorption. We must therefore extrapolate the flux longward of the
Gunn-Peterson trough. But even this continuum is contaminated by
absorption by intervening systems.  Figure~\ref{fig:srhfig3} shows a
``widened spectrum'' of the QSO in order to highlight the absorption
features longward of the Gunn-Peterson trough. (It was constructed by
vertically shifting successive raw spectrograms by 2 pixels and
superposing them.)  There is a break in the spectrum at $\sim
$1620~\AA\,  suggestive of a Lyman limit system at $z=0.773$. We
searched for its \ion{Mg}{2} $\lambda 2800$ absorption but could not
conclusively identify these metal lines, because they are obscured by
strong \ion{H}{1} lines.  There is also a known metal-line system at
$z=1.889$ (Hu et al. 1995). A strong 
absorption line ($W_{\lambda}=1.2$~\AA ) at 1689~\AA\,  
is probably \ion{He}{1} \Lya\,  
$\lambda$584 from this system. Because of the absorption from these
intervening systems, the QSO flux distribution is not a smooth power law 
and hence, not easily extrapolated to shorter wavelengths. 

The lower panel of Figure~\ref{fig:srhfig3} shows the observed flux 
distribution with two continuum estimates superposed. One attempts
to account for the absorption of intervening systems via the program,
CLOUDSPEC (Hubeny 1998), a combination of CLOUDY (\cite{Ferland98}) and 
SYNSPEC (Hubeny et al. 1994). The other (and the one we adopt) is a
flat continuum in wavelength (corresponding to $\alpha = 2$, 
$f_{\nu} \sim \nu^{-\alpha}$) set at
F$_c$=$2.1\times 10^{-16}~{\mathrm erg}~{\mathrm s}^{-1}~{\mathrm
  cm}^{-2}~{\mathrm \AA}^{-1}$. 

\begin{figure}[htp]
  \begin{center}
    \plotfiddle{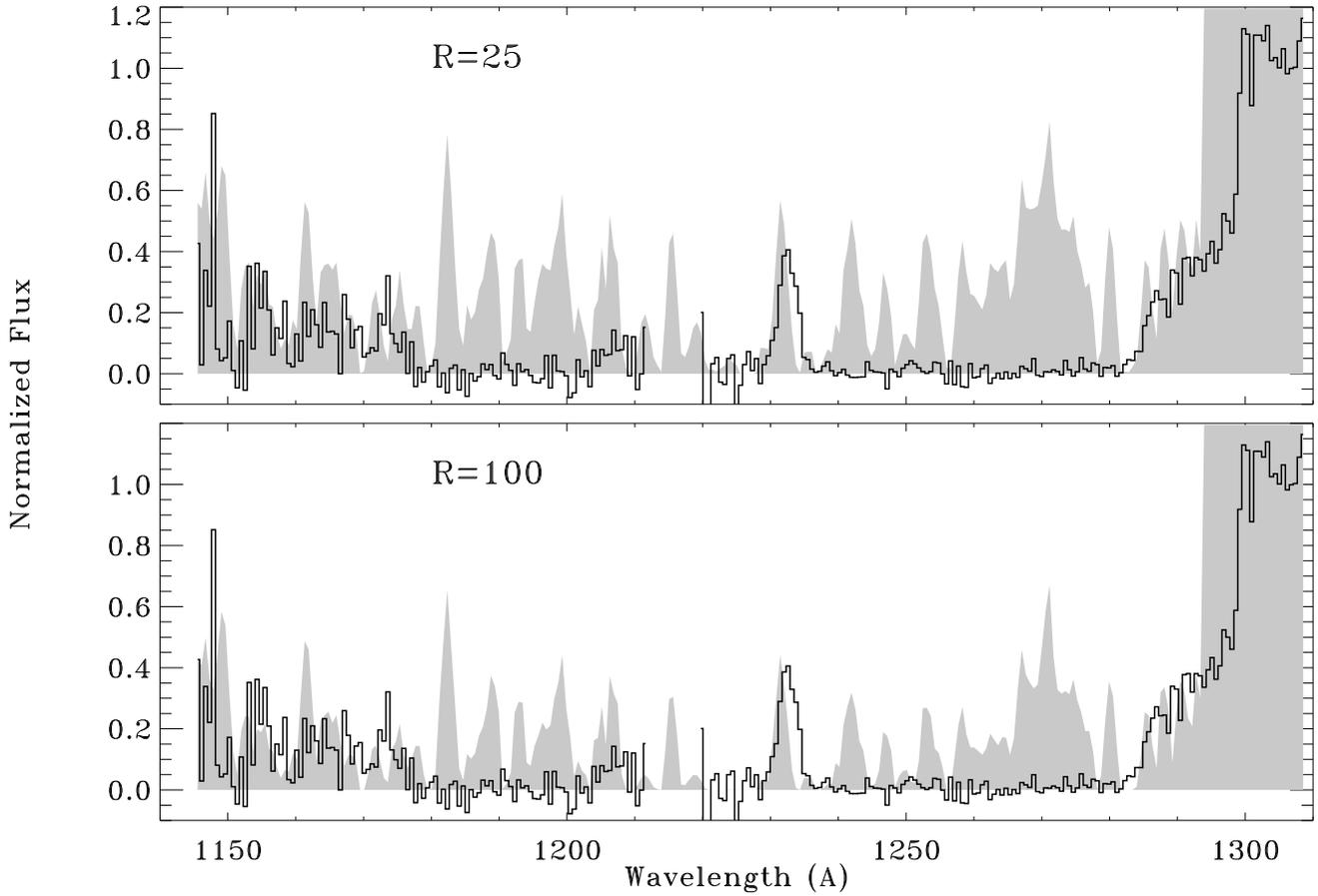}{12cm}{0}{100}{100}{-315}{-335}
    \caption{Comparison of the observed (black line) and
  simulated (filled gray) spectra of Q~0302--003 in the Gunn-Peterson
  trough.  The simulations assumed pure turbulent broadening ($\xi=1$)
  and two different values of the \ion{He}{2}/\ion{H}{1} optical depth
  ratio, R = 25 and 100. The region contaminated by geo-coronal \Lya\, 
  was omitted in order to avoid confusion.\label{fig:srhfig4}}
  \end{center}
\end{figure}

Figure~\ref{fig:srhfig4} compares the STIS spectrum of Q~0302--003
with a simple model of the \ion{He}{2} \Lya\, forest derived from the
high-resolution (6 \kms ) Keck/HIRES spectrum of the \ion{H}{1} \Lya\,
forest. The model calculation involved four steps: (i) estimate the
optical depth in the \ion{H}{1} \Lya\, lines from the flux of each
data-point normalized to the apparent continuum, $\tau_{\mathrm
  HI}(\lambda )=-\ln (I_{\lambda }^{\mathrm HI})$; (ii) calculate the
corresponding optical depth in the \ion{He}{2} lines $\tau _{\mathrm
  HeII}(\lambda )$ at each data-point assuming a ratio of \ion{He}{2}
to \ion{H}{1} optical depths $R = \tau _{\mathrm HeII}(\lambda
)/\tau_{\mathrm HI}(\lambda )$; (iii) reconstitute the \ion{He}{2}
\Lya\, forest, $I(\lambda /4)=\exp (- \tau _{\mathrm HeII}(\lambda
))$; and (iv) degrade the spectrum to the STIS resolution (1.8~\AA ).
The values of $R=25$ and 100 used in the figure correspond to a ratio
of \ion{He}{2} to \ion{H}{1} column densities, $\eta =N_{\mathrm
  HeII}/N_{\mathrm HI}=100$ and 400 respectively, and assume pure
turbulent line broadening. The comparison makes clear what strong
absorptions are part of the \Lya\, forest and what absorption is
associated with the diffuse, or underdense, medium. For example, the
high opacity of \ion{He}{2} \Lya\, in the region, 1260-1280~\AA ,
known as the Dobrzycki-Bechtold void (\cite{Dobrzycki91}; hereafter
the D-B void), must originate in the diffuse medium. On the other
hand, the \ion{H}{1} opacity gap at 1231~\AA\, is also a gap in
\ion{He}{2}. At the very shortest wavelengths, there is a fair
correlation of the observed spectrum with the simulated \ion{He}{2}
\Lya\, forest, suggesting that the opacity of the diffuse medium may
be lower in this region. Close to the QSO redshift, there is also
significant flux that has been interpreted as evidence of the
proximity effect (Hogan et al.  1997).

\subsection{Comparison With Other Lines of Sight}

\begin{figure}[htp]
  \begin{center}
    \plotfiddle{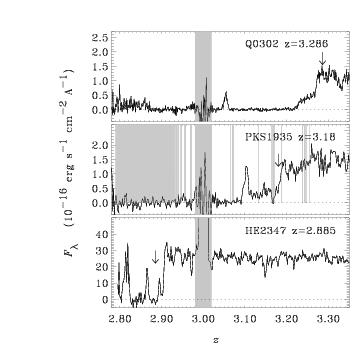}{14cm}{0}{130}{130}{-360}{-110}
    \caption{Comparison of the \ion{He}{2} Gunn-Peterson
  absorption trough plotted as a function of redshift for the
  \ion{He}{2} line at 303.78 \AA.  Q~0302--003 and PKS~1935--692 were
  observed with STIS, and HE~2347-4342 by GHRS. The downward arrows
  denote the redshift of the QSO. As in Figure 4, the spectral region
  near geo-coronal \Lya\, is grayed out.  In the case of PKS~1935--692,
  which has an intervening damped \Lya\, system at $z=0.30$, the
  predicted absorption spectrum from the DLA is also shown in gray as
  a warning that the opacity at $z<3$ should not be interpreted as
  Gunn-Peterson absorption.\label{fig:srhfig5}}
  \end{center}
\end{figure}

To put the line of sight to Q~0302--003 in context,
Figure~\ref{fig:srhfig5} compares the STIS spectrum of Q~0302--003
with those of PKS~1935--692, a $z=3.18$ quasar observed by the GHRS
(Tytler et al. 1995, Jakobsen 1996) and by STIS (Anderson et al.
1999), and HE~2347-4342, a $z=2.885$ QSO observed by GHRS (Reimers et
al. 1997).  (The raw observational data of PKS~1935--692 and
HE~2347-4342 were re-reduced at Goddard.)  The abscissa in this figure
is now the redshift of \ion{He}{2} \Lya , $z=(\lambda/303.78) - 1$,
instead of observed wavelength. In each panel, the downward
pointing arrow shows the wavelength of \ion{He}{2} \Lya\,  at the
redshift of the QSO. Below, we discuss the three major features of the
\ion{He}{2} \Lya\,  absorption trough: the proximity effect, the
absorption trough itself, and gaps in the absorption.

{\it The Proximity Effect.}  This term refers to an observed decrease
in the number of \ion{H}{1} \Lya\, - absorbing clouds in the
neighborhood of a QSO, at least in part caused by the H-ionizing
radiation field of the QSO. A \ion{He}{2} proximity effect involving
both \ion{He}{2}-absorbing clouds and diffuse components of the IGM is
also expected.  It is best defined by the spectrum of Q~0302--003,
although it is also evident in PKS~1935-692; however, note that the
edge of the proximity zone is blocked by a pronounced gap in the
absorption.  In \S\ref{sec:modeling}, we will use the observed flux
distribution in the proximity zone to explore the shape of the
radiation field of the QSO and the metagalactic background at
$z\approx 3$.

{\it The \ion{He}{2} Gunn-Peterson Absorption Trough.}  Strong,
continuous absorption is evident in all three quasar spectra outside
of the proximity zone. Of the three, only the spectrum of Q~0302--003
has the broad baseline in redshift ($z=2.78-3.28$) needed for studying
the evolution of the IGM. The spectrum of PKS~1935--692 would appear
at first also to be useful for this purpose, but it is contaminated by
a $z=0.30$,  damped \Lya\,  system ($\log N_{\mathrm HI} = 21.2$)  whose
Lyman continuum absorption effectively blocks the light from the
quasar corresponding to \ion{He}{2} Ly$\alpha$ at $z < 3.0$ (the
shaded region in Figure 5).  In \S 4, we will analyze the absorption
trough in the spectrum of Q~0302--003 in detail.

{\it Gaps in the Absorption Trough.}  
At wavelengths shortward of the shelf of transmitted flux that we
attribute to the proximity effect, there are
significant gaps in an otherwise solid wall of absorption by \ion{He}{2}. As
can be seen in Figure~\ref{fig:srhfig5}, there are windows of
transmitted flux in the spectrum of Q~0302--003 at $z=3.05$,
PKS~1935--692 at $3.10$, and HE 2347--4342 at $z=2.87$ and $2.82$.
Since these windows of transmitted flux are broader than our
instrumental resolution and are not accompanied by similar, much
narrower ones elsewhere, it is doubtful that they correspond to gaps
appearing by chance among randomly distributed absorbing clouds.
In \S\ref{sec:voidz305}, we shall argue that these coherent stretches of 
transmission are probably the result of
discrete sources photo-ionizing the surrounding material.  

Note how the morphology of low-opacity regions changes with redshift. At
redshifts $z>3$ (and outside of the proximity zone), the opacity gaps
are isolated, discrete, and major. In contrast, regions of transmission
at $z\simlt 2.9$ occur more frequently but with smaller amplitude.
This changing morphology will be discussed in the following three
sections.

\section{THE \ion{He}{2} GUNN-PETERSON ABSORPTION TROUGH}
\label{sec:heiigptrough}

In this section, we make quantitative assessments of the \ion{He}{2} \Lya\,  
optical depth of the IGM along the line of sight to Q~0302--003. These
include both direct determinations of the average transmission within
the \ion{He}{2} Gunn-Peterson absorption trough (\S 4.1), and a more
sensitive but less direct method that makes use of the low level of
fluctuations in transmission in the trough (\S 4.2).  Since the
measurements cover a wide redshift range ($z=2.78-3.28$), we were able
to trace the redshift evolution of the \ion{He}{2} \Lya\,  opacity.  We
find that it rises rapidly with redshift, a result that is compatible
with a shift in the \ion{C}{4}/\ion{Si}{4} ratio between $z=2.9$ and $z=3.0$
reported by Songaila (1998) (\S 4.3). On the assumption that a
lower \ion{He}{2} \Lya\,  opacity is due to a higher fraction of
doubly-ionized helium, we follow the re-ionization history of helium
over the redshift interval, $z=2.8-3.2$ (\S 4.4).

\subsection{Measured Opacities}
\label{sec:measuredopacities}

\begin{figure}[htp]
  \begin{center}
    \plotfiddle{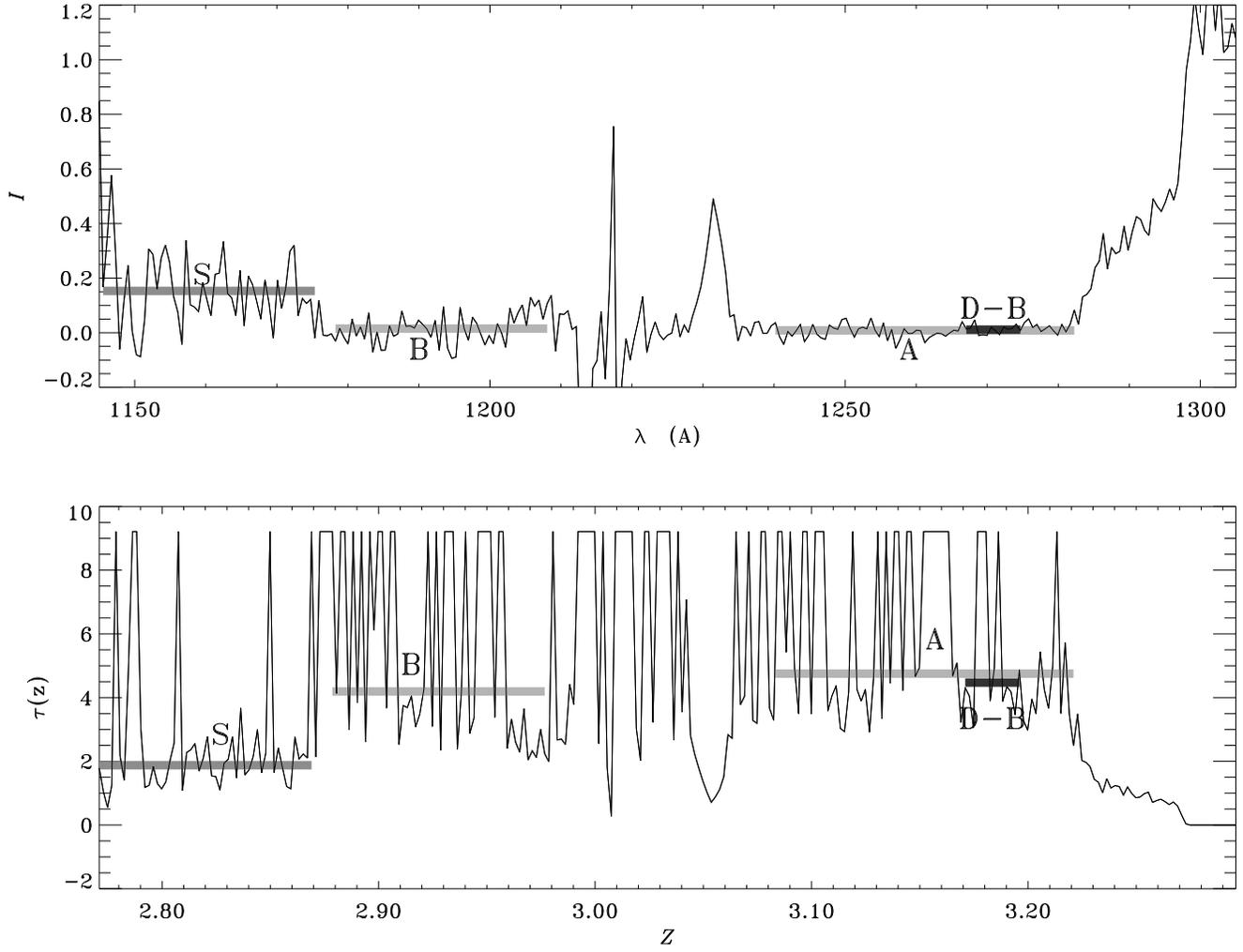}{12cm}{90}{80}{80}{300}{-30}
    \caption{Transmissions (upper panel) and opacities
  (lower panel) in the Gunn-Peterson absorption trough. The selection
  of the different spectral regions is explained in the text. The
  measured values of the apparent transmissions and optical depths are
  given in Table~\ref{tab:gpopacities} and depicted by gray bars.
\label{fig:srhfig6}}
  \end{center}
\end{figure}

We estimated the \ion{He}{2} \Lya\, opacity for different regions as
shown in Figure~\ref{fig:srhfig6}. The upper panel shows the
normalized flux (i.e. the relative transmission of the IGM) as a
function of redshift, while the bottom panel shows the spectrum of
apparent optical depths.  The region labeled ``D-B'' is the
Dobrzycki-Bechtold void (1991).  Regions A and B are used together to
characterize the \ion{He}{2} opacity at $z\approx 3$ for comparison
with theoretical predictions. The region labeled ``S'' corresponds to
redshifts that Songaila (1998) identified as the high-ionization
regime (cf. \S 4.3). We measured the average residual intensities in
each region ($\bar{I}$) and converted them to representative optical
depths ($\tau = -\ln \bar{I}$) with results listed in
Table~\ref{tab:gpopacities}.  Successive columns give the region ID,
the number of data-points in the region, the average transmission and
its uncertainty, the {\it rms\/} error of a single data-point, and the
optical depth corresponding to the average of the transmitted
intensities.

\begin{center}

\begin{table}[ht]

\caption{Measured Opacities in the Gunn-Peterson Trough}
\label{tab:gpopacities}
\begin{tabular}{ccrcrc}
\hline\hline
Region & $\bar{z}$ & \multicolumn{1}{c}{N}  &        \={I}         & $\sigma $(I)&     
$\tau_{\mathrm region}$     \\
\hline                                                       
D-B    &   3.18    &  14                    & 0.0114 $\pm$  0.0044 &   0.0164    & $4.47^{+0.48}_{-0.33}$\\  
A      &   3.15    &  73                    & 0.0086 $\pm $ 0.0029 &   0.0246    & $4.75^{+0.41}_{-0.29}$\\
B      &   2.93    &  52                    & 0.0151 $\pm $ 0.0079 &   0.0567    & $4.19^{+0.73}_{-0.42}$\\
S      &   2.82    &  52                    & 0.1532 $\pm $ 0.0179 &   0.1292    & $1.88^{+0.12}_{-0.11}$\\
A+B    &   3.06    & 125                    & 0.0113 $\pm $ 0.0037 &   0.0410    & $4.48^{+0.39}_{-0.28}$\\  
\hline
\end{tabular}
\end{table}
\end{center}

In agreement with previous GHRS
observations, the optical depth of the D-B void $\tau_{\mathrm {DB}} =
4.47^{+0.48}_{-0.33}$ is quite high, implying that the absorption
appears to be due to diffuse or underdense material.  Alternatively,
there may be a nearby starburst galaxy or other soft ionizing source 
that could ionize \ion{H}{1} but not \ion{He}{2}, thereby producing a low
\ion{H}{1} opacity but not affecting the \ion{He}{2} opacity.  As the
GHRS spectrum only sampled the Gunn-Peterson trough above $z=3.1$, the
measured opacities in the other regions (A, B, S) are new, and will
be discussed later in this section.  


The $rms$ errors quoted in the table give a measure of the random
errors.  There remains the question of systematic errors, which are
the more serious of the two. Since the STIS detector is a
pulse-counting detector with a linear response at low light levels, we
can rule out the possibility that the low residual flux in the
Gunn-Peterson trough is due to a systematic error in the response
function. However, there could be a systematic error in the residual
flux if the cross-dispersion (``vertical'') gradation in the
background were not linear, so that the average of the upper and lower
background regions would not be representative of the background under
the QSO spectrum. We can put an upper limit on this systematic error
by re-reducing the spectrograms with the upper background or lower
background only. These alternate reductions produce residual fluxes in
the Gunn-Peterson absorption trough that are well within the
uncertainties quoted in the table. Other reduction procedures
(15-point background smoothing applied twice, registration of the
observations before co-addition) yield differences on the same order
as adopting the upper or lower background only. We therefore interpret
the errors listed in the table as representing total errors.

\subsection{Opacity Fluctuations}
\label{sec:opacityfluctuations}

Now that we have explicitly measured the opacities, we can obtain a
more sensitive, but less direct result by looking for gaps in the
opacity that appear by chance and comparing them with predictions.  We
find that a high opacity is required to produce the broad stretches of
nearly zero transmission as seen, for example, in spectral region A.
This region has 24 independent samples of width $\Delta z=0.0058$,
where independent samples are assumed to be separated by the
instrumental resolution (3 pixels).  All 24 samples show a relative
transmission, $\bar{I}\leq 0.030$ (cf. Figure 2), implying that the
optical depths exceed $-\ln(0.03)=3.50$ everywhere in this region. We
designate this opacity as $\tau_{\mathrm limit}$.  Even in the absence
of extra ionizing sources, we should expect to see some variations
caused by the random gaps in the overlapping of clouds.  The fact that
such variations do not cause observable transmission spikes implies
that the average optical depth is greater than a certain amount -- a
value that we shall now evaluate.

Fardal, Giroux \& Shull 
(1998) constructed a model for the clouds in the IGM based on the
observed properties of the Ly$\alpha$ forest and then, on the basis of
Monte Carlo simulations, found that fluctuations in the {\it apparent
  opacity\/} of \ion{He}{2}, defined as $\tau\equiv - \ln \bar{I}$,
over an interval $\Delta z$, follow the approximate relationship
(their Eq.~17):
\begin{equation}
\label{tau_fluc}
  \left( 
    {\Delta \tau}\over \bar{\tau} 
  \right)_{\mathrm model}
  \approx 
  0.03 
  (\xi\, \eta_{100})^{-0.1} 
  \Delta z^{-0.5} 
  \left(
      {1+z}\over 4
  \right)^{-1.0}~, 
\end{equation}
where $\bar{\tau}$ is the average of $\tau$ over
many intervals of $\Delta z$, and $\Delta\tau$ represents a deviation
of $\tau$ from this average.  The quantity 
$\xi$ is the characteristic ratio of the \ion{He}{2} line width
to that of \ion{H}{1} (ranges from 0.5 to 1.0, depending on whether the
line broadening is dominated by thermal or turbulent processes), and
$\eta_{100} = N_{\mathrm HeII}/(100 ~ N_{\mathrm HI})$. 


Now consider the (very small) probability $p$ that a single sample of
width $\Delta z$ shows a spike $\bar{I}\geq 0.030$, i.e. $\tau_{\mathrm
  limit} \leq 3.50$.  This quantity $p$ then represents the one-tail
area below $\tau_{\mathrm limit}$ for a normal distribution with a
standard deviation $\Delta \tau$ centered on $\bar{\tau}$.
The chances of not seeing a spike over all 24 samples is $e^{-24~p}$.
We now determine the limiting condition that would create a 10\%
expectation of seeing no spikes in any of the samples, since this
leads to a confidence level of 90\% that we are not overestimating
$\bar{\tau}$ and/or underestimating $\Delta \tau$.  This happens when
$p=-\ln(0.1)/24=0.096$ which is the one-tail area for $\sigma=1.30$
standard deviations below the center of the normal distribution.  This
requirement sets a lower limit for the average opacity,
\begin{equation}
\label{tau_result}
\bar{\tau} > \tau_{\mathrm limit} 
       \left[ 
         1-\sigma \left( {\Delta \tau \over \bar{\tau}}\right)_{\mathrm model}
       \right]^{-1}~.
\end{equation}
For $\tau_{\mathrm limit}=3.50$, $\sigma=1.30$, $\eta_{100}$ =3.5
(cf. \S\ref{sec:etafarfromqso}), and $\xi = 0.5$ or $1.0$, we obtain $\bar{\tau}
> 6.6$ and $6.2$, respectively. Both values are somewhat higher than the 
measured value, $\tau_{\mathrm A}=4.75\pm 0.4$ in Region A. 

\subsection{A Break in \ion{He}{2} Ly{\boldmath $\alpha$} 
Opacity at {\boldmath $z=2.9-3.0$}?}
\label{sec:opacitybreak}

Songaila (1998) recently reported on a study of Keck HIRES spectra of
13 quasars, in which she found that the \ion{C}{4}/\ion{Si}{4} ratio
at $z = 2.9$ was 3.4 times higher than at $z = 3.0$. Since the
ionization potentials of \ion{Si}{4} (45.1~eV) and \ion{C}{4}
(64.5~eV) straddle the ionization potential of \ion{He}{2} (54.4~eV),
\ion{C}{4}/\ion{Si}{4} serves as a proxy for the ionization state of
helium, and she attributed the change in this ratio to a hardening of
the metagalactic spectrum below $z=3.0$. However, Songaila's findings
were disputed by Boksenberg et al. (1998) who found no sudden change
in ionization, although they did note a gradual increase in
\ion{C}{4}/\ion{Si}{4} toward lower redshifts between $z\approx 3.8$
and $z\approx 2.2$.

The STIS spectrum of Q~0302--003 provides an opportunity to trace the 
ionization evolution of helium over a broad redshift range,
$z=2.8 - 3.2$.  (We assume that a lower \ion{He}{2}
opacity means a higher fraction of doubly ionized helium.) Our
calculations with CLOUDY (\cite{Ferland98}) indicate that the
\ion{He}{3}/\ion{He}{2} ratio should change by at least as much as
\ion{C}{4}/\ion{Si}{4} for a Madau et al. (1999) background radiation
field at $z=3$. We therefore looked to see whether the \ion{He}{2}
opacities show a distinct change between $z=2.9$ and $z=3.0$. We find
that Region S of Figure~\ref{fig:srhfig6} does indeed show a residual
flux in the Gunn-Peterson trough at $z<2.87$ implying an average
transmission that corresponds to $\tau_{\mathrm S}= 1.88$ in this region.

The redshift interval, $z=2.77-2.87$, however, is by no means a void
in the conventional sense.  Many \ion{H}{1}  \Lya\,   forest lines occur
there, in some cases with clear corresponding increases in \ion{He}{2} 
opacity. To compare the $z<2.9$ and $z>3.0$ opacities, we selected
those data points in Region S unassociated with features in the Lyman
forest, i.e., positions identified as having a normalized \ion{H}{1}
transmitted flux greater than 0.95 even when degraded to STIS
resolution.  We find 10 non-contiguous data-points in Region S
fulfilling this $I_{\mathrm HI}>0.95$ requirement.  The  \ion{He}{2} 
opacity corresponding to the average transmission for these
10 data-points is $\tau =1.41 \pm 0.06$, which is a factor of 3
times lower than in the D-B void. Similarly, the opacity for the whole
Songaila region ($\bar{z}=2.82$) which includes numerous \Lya\,   lines,
$\tau_{\mathrm S}=1.88$, is a factor of 2.5 times lower than for Region A at
$\bar{z}=3.15$. We conclude that the \ion{He}{2}
Gunn-Peterson data are in accord with Songaila's assertion of an
abrupt shift in ionization between $z=2.9$ and $z=3.0$.


\subsection{Comparison with Cosmological Models}

Observations of  
\ion{He}{2} \Lya\,  absorption give  a direct connection with
cosmological simulations if differences in resolution are taken into 
account. For example, simulations  
by Zhang et al. (1998) cover a cube only 9.6 comoving Mpc on a side
with a cell size of 75 kpc$^{3}$.  In contrast, our STIS spectrum of
the G-P trough covers a much larger distance (400 comoving Mpc) but
at a coarser resolution ($\sim 2$ comoving Mpc), so
it smooths over all of the fine-scale fluctuations predicted by
theory. Nevertheless, we can apply the same kind of smoothing on the
theoretical opacities by taking averages.  
Zhang et al. (1998, their figure 20) predict that the average
\ion{He}{2} opacity should increase from $\bar{\tau}=0.50$ at $z=2.4$, to
$\bar{\tau}=0.83$ at $z=3$, and $\bar{\tau}=2.43$ at $z=4$.  

\begin{figure}[htbp]
  \begin{center}
    \plotfiddle{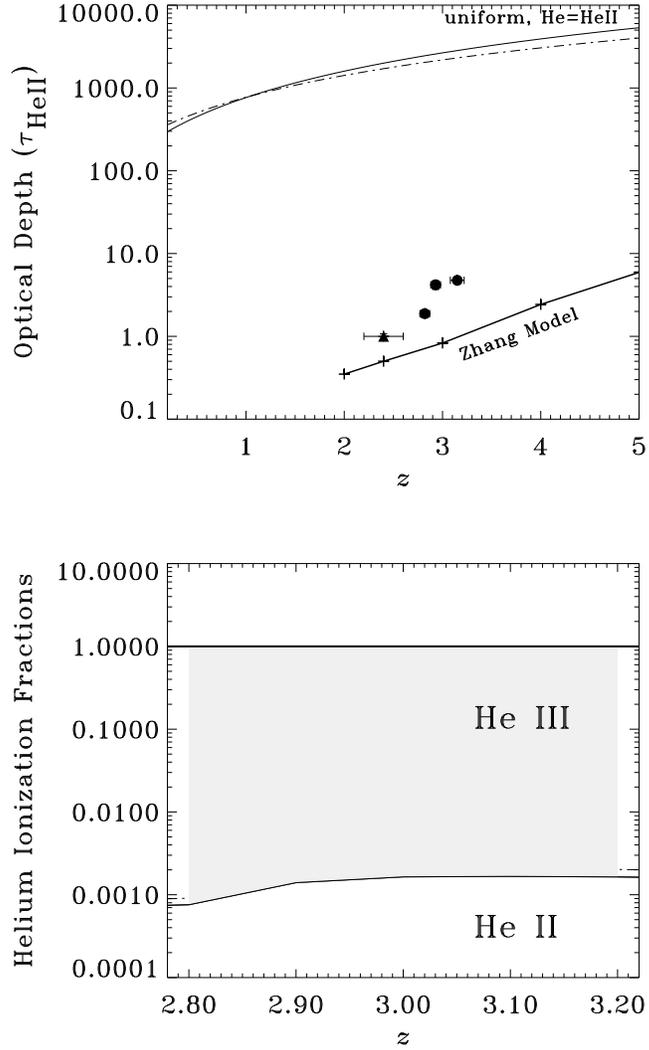}{12cm}{0}{65}{65}{-200}{-60}
    \caption{Evolution of the \ion{He}{2} \Lya\, opacity
  (top) and helium ionization fractions (bottom).  Top: The filled
  circles represent data from our STIS observations of Q~0302--003,
  while the filled triangle represents the HUT results for HS
  1700+6416 (Davidsen et al.  1996). The bold curve shows the
  opacity predicted by Zhang et al. (1998). The other curves show the expected
  \ion{He}{2} \Lya\, opacity if all helium were singly ionized and
  uniformly distributed.  The assumed cosmological parameters are:
  $H_{\mathrm o}$=65 \kms Mpc$^{-1}$ and $\Omega _{\mathrm o}=0.2$,
  $\Lambda=0$ (solid line) , or $H_{\mathrm o}$=50 \kms Mpc$^{-1}$ and
  $\Omega _{\mathrm o}=1.0$, $\Lambda=0$ (dash-dot).  Bottom: Implied
  helium ionization fractions in the redshift range observed by STIS,
  based on the assumption that all helium is once or doubly ionized.
  The fraction of HeII is only about 0.1 \% at $z\sim 3$.
\label{fig:ev_tau}}
  \end{center}
\end{figure}

The measurements listed in Table 2 imply that the total \ion{He}{2}
opacity rises more rapidly with redshift than the models (Zhang et al. 1998, 
Fardal et al. 1998).  As shown in Figure \ref{fig:ev_tau}, which combines
data for HS~1700+6416 (\cite{Davidsen96}) with the measured opacities
derived from Regions S, B, and A in the spectrum of Q~0302--003, the 
\ion{He}{2} optical depth in \Lya\, increases sharply from $\tau = 1.0$
at $z=2.4$ to $\tau \sim 4.7$ at $z=3.15$.  The more sensitive
estimates of the \ion{He}{2} opacity based on the lack of opacity
fluctuations suggest that it may rise with redshift even faster than
shown.  The top panel also plots the mean \ion{He}{2} \Lya\ opacity predicted 
by Zhang et al.'s (1998) cosmological model. It shows that the model 
generally underestimates the \ion{He}{2} \lya\ opacity and 
does not reproduce the
observed steep rise in opacity with increasing redshift. 

The top panel of Fig. \ref{fig:ev_tau} also shows the opacity expected
if all helium were in the form of He$^+$ and were uniformly
distributed.  Two curves are shown: one for our assumed cosmological
parameters (solid line), the other for Madau et al.'s (1999), i.e.
$H_{\mathrm o}$=50 \kms Mpc$^{-1}$, $\Omega _{\mathrm o}=1.0$ and
$\Lambda=0$. In either case, the measured opacity is more than 600
times smaller than expected.  The fact that we detect a small residual
intensity in Regions A and B indicates that most, but not all, of the
helium is doubly ionized at $z=3$.  Accordingly, the bottom panel
shows the implied fractions of \ion{He}{3} and \ion{He}{2}.

In summary, away from distinct gaps in the \ion{He}{2} absorption,
the \Lya\,  opacity shows a sharp rise between $z=2.8$ and $z=3.2$.
At $z\sim 3.15$, the optical depth is estimated at 4.7 or more. It
therefore appears that we are witnessing the transition from a 
\ion{He}{2} \Lya\,  forest to a \ion{He}{2} Gunn-Peterson trough.
The sharp rise in opacity at $z\sim 3$ (interpreted as an increased
fraction of \ion{He}{2} ) is consistent with Songaila's claim of a lower
ionization level of the IGM at $z> 3$ . However, we caution that 
these  results are based on a single line of sight. Other lines of sight
may give different results since \ion{He}{2} is such a minor fraction 
of helium (0.16\%). In fact, some of the large scatter in 
Songaila's measured \ion{C}{4}/\ion{Si}{4} ratios may reflect such 
differences. We plan to pursue the question of a ``universal epoch
of reionization'' by studies of the \ion{He}{2} Gunn-Peterson effect 
along the sightlines to PKS~1935--692 and HE~2347--4342.

\section{MODELS OF GUNN-PETERSON ABSORPTION AND THE PROXIMITY EFFECT}
\label{sec:modeling}

The \ion{He}{2} opacities measured in
\S\ref{sec:heiigptrough} imply that the vast majority of He is
doubly ionized at $z\sim 3$, presumably by the UV background.
In this section, we derive the properties of the UV background at
$z\sim 3$ from the \ion{H}{1} and \ion{He}{2} \Lya\,
spectra of Q~0302--003. To pursue the matter, we constructed
spectral models that take into account just one ionizing source,
the QSO itself, in addition to the UV background. These models allow
us (1) to identify regions within the proximity zone that
are particularly sensitive to the shape of the UV background, and can
thus be used to constrain it, and (2) to discriminate among the 
\ion{He}{2} opacity gaps those likely caused by underdense regions in the
IGM from those due to discrete ionizing sources. 
In the following
sections, we describe the model input, calculations, and results.

\subsection{Model Input} 

The three main inputs to the models are the observed flux distribution
in the \ion{He}{2} Gunn-Peterson absorption trough, the QSO flux
distribution, and the physical parameters of the \ion{H}{1} \Lya\,  
forest. Below, we describe our methods for estimating the QSO
continuum flux distribution and the physical characteristics of the
\ion{H}{1} \Lya\,  lines.

\subsubsection{The spectral energy distribution of Q~0302--003}
\label{sec:qso_spectrum}

\begin{figure}[htbp]
  \begin{center}
    \plotfiddle{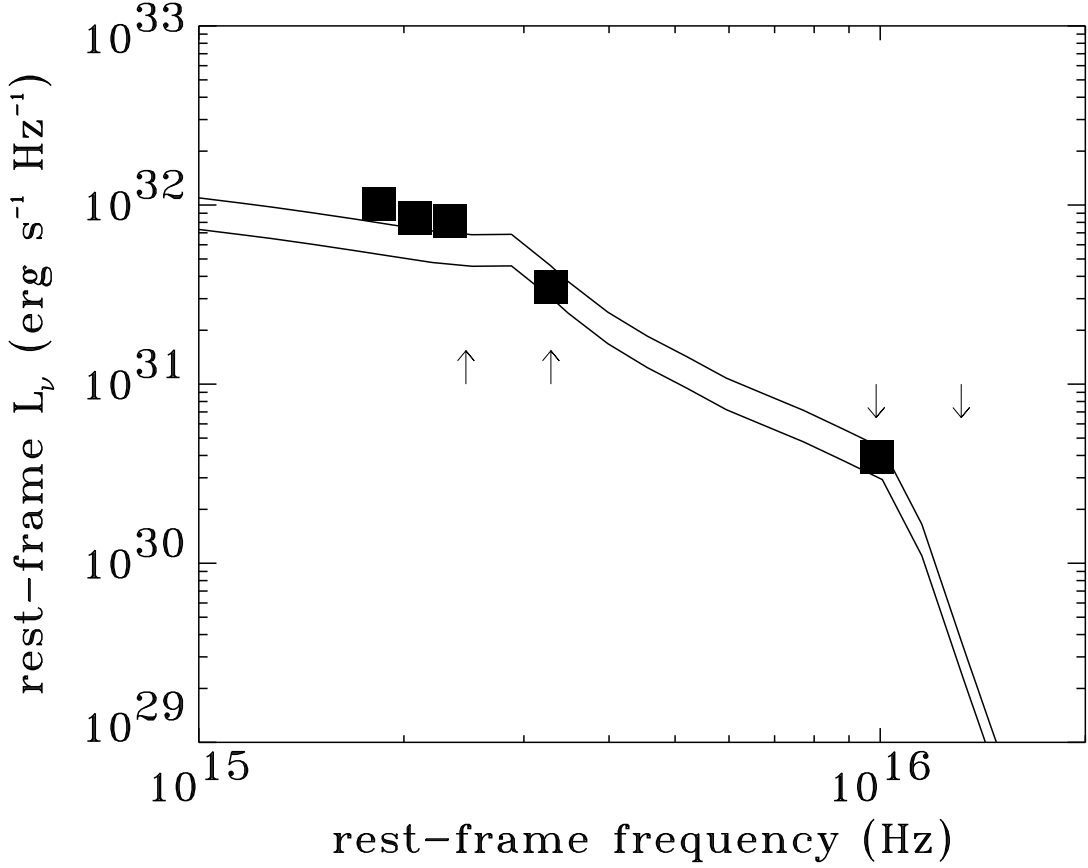}{12cm}{0}{100}{100}{-340}{-345}
    \caption{Probable spectral energy distribution of Q~0302--003.
  The filled squares show the observed continuum flux transformed to
  the quasar rest frame. The upward-pointing arrows give the
  frequencies of HI \Lya\, (left) and the HI Lyman limit (right). The
  downward-pointing arrows show the same quantities for \ion{He}{2}.
  The solid lines show two theoretical models of a bare accretion disc
  around a super-massive, Kerr black hole with a black-hole mass $M= 2
  \times 10^{10}$ $M_\odot$ and the mass accretion rate $\dot M = 24$
  $M_\odot \, {\mathrm yr}^{-1}$ (upper curve); and $M= 1.6 \times
  10^{10}$ $M_\odot$ and mass accretion rate $\dot M = 16$ $M_\odot \,
  {\mathrm yr}^{-1}$ (lower curve).  Both models assume an almost
  face-on disk, with inclination $\cos i = 0.99$.
\label{fig:hubenyfig1}}
  \end{center}
\end{figure}

We estimated the ionizing flux distribution of the QSO from
measurements of the apparent continuum combined with model continuum
flux distributions.  Figure~\ref{fig:hubenyfig1} plots the theoretical
luminosity distribution of Q~0302--003 along with observationally
derived continuum points transformed to the quasar rest frame
($z=3.29$).  The observed points are taken from our STIS data and from
the ground-based spectrum obtained by Sargent, Steidel, \& Boksenberg
(1989). The STIS observations cover the restframe interval, 267 -
402~\AA , or $7.5\times 10^{15} - 1.12\times 10^{16}$ Hz.  The SSB
fluxes were multiplied by a factor of 1.25 to account for the
depression of the apparent continuum by the multitude of lines that
make up the \Lya\, forest. This factor was estimated from the
normalized Songaila et al. spectrum (Table 1) degraded to the
resolution of the spectrum obtained by Sargent et al.  We found that a
power law of the form $f^{\rm QSO} = f_{\mathrm HI}^{\mathrm QSO}~
(\nu/\nu_{\mathrm HI})^{-\alpha_{\mathrm HI}^{\rm QSO}}$ with
${\alpha_{\mathrm HI}^{\mathrm QSO}} = 1.9$, gives a good
representation of the data from the optical to the far-UV
($\lambda_{\mathrm rest}= 1000 - 304$~\AA).  In our models of the QSO
flux distribution, we extrapolate this power law into the unobserved
\ion{He}{2} Lyman continuum as well.

The two theoretical models shown in the figure are for a bare
accretion disk around a super-massive, Kerr black hole, one with a
black-hole mass $M= 2 \times 10^{10}$ $M_\odot$, and accretion rate
$\dot M = 24$ $M_\odot \, {\rm yr}^{-1}$ (upper curve); and the other
with $M= 1.6 \times 10^{10}$ $M_\odot$ and mass accretion rate $\dot M
= 16$ $M_\odot \, {\rm yr}^{-1}$ (lower curve).  Both models assume
the maximum stable rotation of the black hole, with a specific angular
momentum $(a/M) = 0.998$.  The integrated spectrum of the disk taking
into account all general relativistic effects was computed by program
KERRTRANS (\cite{Agol97}); the best fit was obtained for disk seen
almost face-on. A detailed description of the modeling procedure is
given by Hubeny \& Hubeny (1998); the presented models are taken from
the grid of Hubeny, Blaes \& Agol (1998).  As these spectra represent
a bare accretion disk, they do not take into account the effects of a
Comptonizing corona.

The theoretical flux distribution has three distinct spectral regions,
each with its characteristic spectral slope, $\alpha$, defined by
$f \propto \nu^{-\alpha}$. For frequencies $\nu <$ $3
\times 10^{15}$ Hz ($\lambda_{rest} > 1000$~\AA), $\alpha \approx
0.55$.  For $\nu = 3 \times 10^{15} - 1\times 10^{16}$ Hz
($\lambda_{rest}= 1000 - 300$~\AA), $\alpha \approx 2$.  At $\nu >
10^{16}$ Hz, the flux falls off steeply, and $\alpha \approx 11$.  The
models reproduce the observed flux in the first two regions. In the
high-frequency region, where we have no observed data, the theoretical
flux is likely a lower limit, because the models do not take into
account the effects of a Comptonizing corona, which increase the flux
dramatically.  Therefore, we conclude that the power-law slope of the 
\ion{He}{2} Lyman continuum in the QSO spectrum is $ 2 \la
\alpha_{\mathrm  HeII}^{\mathrm QSO} \la 11 $.

\subsubsection{Line lists}
\label{sec:linelists}

In order to estimate the value of $\eta= N_{\mathrm HeII}/N_{\mathrm
  HI}$ as a function of $z$, we made use of a line list containing the
observed \Lya\, lines whose parameters ($z, \NHI ,b_{\mathrm HI}$)
were derived from VPFIT measurements of the Keck HIRES spectra (cf.
\S\ref{sec:keckhiresspectra}).  This line list is hereafter referred
to as the `observed line list'.  Following Songaila et al. (1995), we
expanded the sample in the range of $9 < \log{(N_{\mathrm HI})} < 13$,
by producing a line list with random values of \NHI , $b$ and $z$,
whose distribution parameters were obtained by interpolating in
redshift the values determined by Kim et al. (1997).  The resulting
line list is called the `combined line list'. Note that even if $\eta
= 1000$, the $\log{N_{\mathrm HI}} = 9$ clouds do not contribute much
to the \ion{He}{2} opacity.

\begin{figure}[htp]
  \begin{center}
    \plotfiddle{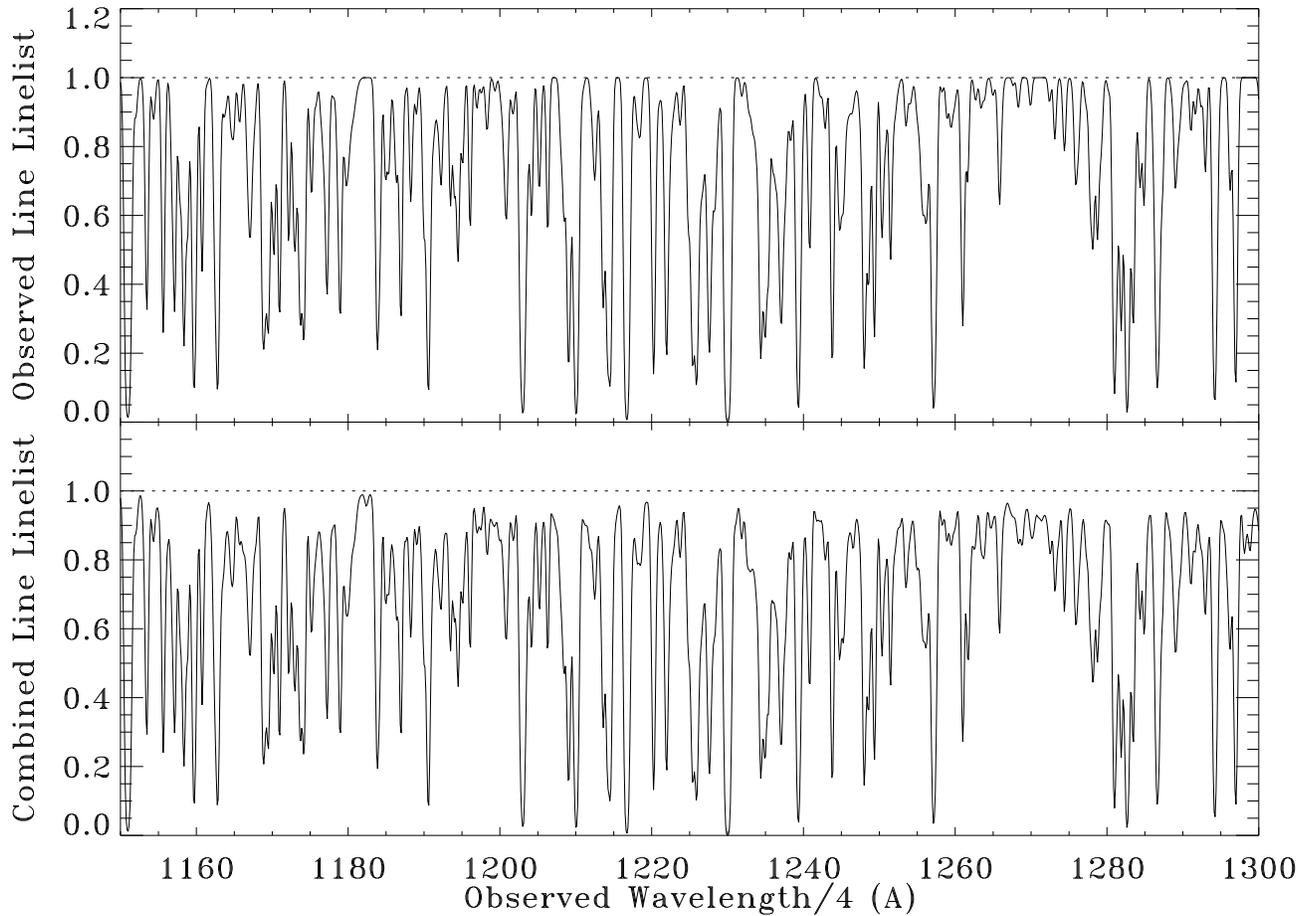}{14cm}{0}{100}{100}{-315}{-330}
    \caption{Effect of very weak \Lya\, lines on the observed
  \ion{H}{1} \Lya\,  forest. Top: the normalized spectrum of the \Lya\, 
  forest computed with observationally derived parameters, \NHI , $b$
  and $z$. Bottom: the same as above except that the ``combined'' line
  list was used (see text). The effect of including very weak lines is
  to produce a shallow \ion{H}{1} Gunn-Peterson trough by
  depressing the apparent continuum by about 6\%.
\label{fig:comb_hi}}
  \end{center}
\end{figure}

The immediate effect of including this plethora of very weak lines is
to depress the apparent continuum in the \ion{H}{1} \lya\ forest, 
i.e. to produce a shallow
\ion{H}{1} Gunn-Peterson absorption trough.  Figure~\ref{fig:comb_hi}
compares the \ion{H}{1} \Lya\,  forest computed with the combined line
list vs. the observed line list. The predicted Gunn-Peterson trough
has a depth of about 6\%, which is shallow enough to escape detection
in real data. However, it is consistent with Fang et al.'s (1998)
measurement ($\tau_{\mathrm HI}=0.113 \pm 0.02$) of the \ion{H}{1}
Gunn-Peterson effect in the $z=3.787$ QSO, PKS 1937--101.

\subsection{Model Calculations}
\label{sec:n_heii}

The model calculation is recursive.  First, we consider the cloud    
closest to the QSO: we assume that it sees the `pure' QSO spectrum and
the UV background. We compute its \ion{H}{1} and \ion{He}{2}
photo-ionization rates, whose ratio is proportional
to the value of $\eta$ for this cloud (with the simplifying
assumptions used), and allows the derivation of its
\ion{He}{2} column density. We then consider the second closest cloud
to the QSO: it sees the UV background and the QSO whose spectrum
shortward of the \ion{H}{1} and \ion{He}{2} Lyman limit is depressed  
by the continuum absorption  due to the first cloud. Since we know
the \ion{H}{1} and \ion{He}{2} column densities and redshift of the
first cloud, we can compute the QSO spectrum seen by  the second cloud, the
resulting  \ion{H}{1} and \ion{He}{2} photo-ionization rates; and
from the observed \ion{H}{1} column density of the second cloud, we 
compute its \ion{He}{2} column density. 
The process is then repeated for each cloud of the
considered line list. We then use the calculated value of $N_{\mathrm
  HeII}$ and the line-width ratio $b_{\mathrm   HeII}/b_{\mathrm
  HI}=1$ (pure turbulent broadening) to calculate the \ion{He}{2}
absorption spectrum (Voigt profiles) of all the lines in the line list. 

The main difference between the development made here and the theory
of Str{\"o}mgren spheres is the presence of a diffuse ionizing
radiation field (the UV background) which leads to a much more gradual
stratification in ionization compared to the sharp edge of a Str{\"o}mgren
sphere. Also, contrary to a Str{\"o}mgren sphere, the gas is
considered to be inhomogeneous, i.e. composed of numerous clouds embedded
in a diffuse medium. Finally, there is a velocity gradient  within the 
ionized region (due to the expansion of the Universe) so that clouds far
from the main ionizing source, i.e. the QSO, recede faster from the
QSO than do closer ones.

\subsubsection{Assumptions}

To compute the ratio of column densities of \ion{He}{2} and
\ion{H}{1}, we assume that \Lya\, clouds at $z \sim 3$ are in
photo-ionization equilibrium with hydrogen mostly ionized and helium
mostly doubly ionized.  Indeed, since all the \Lya\, clouds in front of
Q~0302--003 have $\log{N_{\mathrm HI}} \leq 15.47$, most \ion{H}{0} is
certainly ionized, and Figure~\ref{fig:ev_tau} shows that nearly all
of the helium must be doubly ionized.  We also assume that the
ratio of column densities of \ion{He}{2} and \ion{H}{1} is given by
\begin{equation}
  \label{eq:n_heii__n_hi}
  \eta \equiv
  \frac{N_{\mathrm HeII}}{N_{\mathrm HI}} = 
  \frac{n_{\mathrm He}}{n_{\mathrm H}}~
  \frac{\alpha_{\mathrm HeII}}{\alpha_{\mathrm HI}}~
  \frac{\Gamma_{\mathrm HI}}{\Gamma_{\mathrm HeII}},
\end{equation}
where $\alpha_i$ is the recombination coefficient of ion $i$
(\ion{H}{1} or \ion{He}{2}), $\Gamma_i$ is its ionization rate, and
$n_{\mathrm He}$, $n_{\mathrm H}$ are the total number densities of
helium and hydrogen respectively. Note that for clouds with high
\NHEII , self-shielding and emissivity of the clouds start  to be
substantial, and this simple equation is no longer valid. 
For $n_{\mathrm He}/n_{\mathrm H} = 0.082$ and
$\alpha_{\mathrm HeII}/\alpha_{\mathrm HI} =5.418 $ (cf.  Osterbrock
1974), the equation reduces to 
\begin{equation}
\label{eq:eta_44}
\eta = 0.44 \, \frac{\Gamma_{\mathrm  HI}}{\Gamma_{\mathrm HeII}}.  
\end{equation}
The photo-ionization rates are given
by:
\begin{equation}
  \label{eq:ionization_rate}
  \Gamma_i = \int_{\nu_i}^\infty~
  \frac{f}{h \nu}~\sigma_\nu~{\rm d}\nu,
\end{equation}
where $f$ is the total (QSO + UV background) 
ionizing photon flux, $\sigma_\nu \simeq \sigma_i ~(\nu/\nu_i)^{-3}$ 
is the  photo-ionization cross-section (cf. for example, Osterbrock
1974), and $\nu_i$ is the frequency of  ion $i$ Lyman limit.
At all redshifts,  the photo-ionization rates due to the UV background, 
$\Gamma_i^{\mathrm J}$,  are related  to the intensities at the
Lyman limits $J_i$ by the relation: 
\begin{equation}
  \label{eq:photo_ionization_rate_J}
  \Gamma_i^{\mathrm J} = 
  \frac{4 \pi \sigma_i}{h}~ 
  \frac{J_i}{3+\beta_i^{\mathrm J}},
\end{equation}
if the Lyman continuum radiation of the UV background spectrum can be
represented by a power-law $J(\nu) = J_i
(\nu/\nu_i)^{-\beta_i^{\mathrm J}}$.  Useful,
numerical relations between these quantities are: 
\begin{eqnarray}
  \label{eq:gamma_J}
\Gamma_{\mathrm HI}^{\mathrm J}   & = & 2.99 \times 10^{-12}~ 
                                        \frac{J_{\mathrm HI}}{10^{-21}} ~
                                        \frac{4}{3+\beta_{\mathrm HI}^{J}}~
                                        {\mathrm s}^{-1}, \nonumber\\
\Gamma_{\mathrm HeII}^{\mathrm J} & = & 7.47 \times 10^{-15}~ 
                                        \frac{J_{\mathrm HeII}}{10^{-23}} ~
                                        \frac{4}{3+\beta_{\mathrm HeII}^{J}}~
                                        {\mathrm s}^{-1}.
\end{eqnarray}

\subsubsection{The first cloud}

Starting with the closest cloud to the quasar and moving towards
decreasing redshifts,
we compute the expected value of $\eta$ as the
contributions from the quasar decrease while the UV background remains
the same. The photo-ionization rates of the first cloud at
$z = z_{\mathrm 1}$ are:
\begin{equation}
  \label{eq:1st_cloud}
  \Gamma_i^{\rm 1} = 
  \Gamma_i^{\mathrm J}
  +
  \frac{\sigma_i}{h} ~
    \frac{f_i^{\rm QSO}}{3+\alpha_i^{\rm QSO}}  ~
    \left(
      \frac{1+z_{\mathrm QSO}}{1+z_1} ~
    \right)^{-\alpha_i^{\rm QSO}},
\end{equation}
where the QSO flux at a redshift $z$ and frequency $\nu \ge \nu_i$
is $f^{\rm QSO}(z,\nu) = 
f_i^{\rm QSO}(z)~[(1+z_{\mathrm QSO})~\nu/\nu_i]^{-\alpha_i^{\rm QSO}}$.
The quasar flux at the Lyman limit seen by a cloud at redshift $z$ 
is related to the QSO luminosity 
$L_i^{\rm QSO}$ by: 
\begin{equation}
\label{eq:inverse_square_law}
f_i^{\rm QSO}(z) = \frac{1+z}{1+z_{\mathrm QSO}}~
  \frac{L^{\rm QSO}}{4 \pi D_{\mathrm L}^2 }, 
\end{equation}  
where $D_{\mathrm L}(z,z_{\mathrm QSO})$ is the luminosity distance between 
the cloud  and the QSO (cf.  Kayser, Helbig \& Schramm 1997). 
The photo-ionization rates of \ion{H}{1} and \ion{He}{2} 
are calculated with Eq. \ref{eq:1st_cloud}, and then 
Eq. \ref{eq:eta_44} is used to compute $\eta$.  The value of the
\ion{He}{2} column density  is then easily calculated from the observed
\ion{H}{1} column density.

\subsubsection{The ${\mathrm k}$th cloud}

Since the QSO spectrum below the ion $i$ Lyman limit seen by the $k$th
cloud ($k \ge 2$) is depressed by the continuum opacity produced by
the $k-1$ clouds located between it and the QSO, the photo-ionization
rates of the $k$th cloud at $z = z_{\mathrm k}$ are:
\begin{equation}
  \label{eq:l_cloud}
  \Gamma_i^{\rm k} = 
    \Gamma_i^{\mathrm J}
  +
  \int_{\nu_i/(1+z_{\mathrm k})}^\infty~
  \frac{f}{h~ \nu}~\sigma_\nu~
  \exp{\left(-\sum_{l=1}^{k-1}~ \tau_i^{\rm l}\right)}~{\rm d}\nu 
\end{equation}
where 
$\tau_i^{\rm l} = N_i^{\rm l}~\sigma_i~(\nu (1+z_{\mathrm l})/\nu_i)^{-3}$.
The values of $N_{\mathrm HI}^{\rm l}$ are obtained from the observations and
the values of $N_{\mathrm HeII}^{\rm l}$ have been estimated at the
calculation of the  $l^{\mathrm th}$ cloud.
Expanding the exponential as a series 
($\exp{(-x)} = \sum_{\mathrm m=0}^{\infty} (-x)^m/m!$) 
we obtain an analytical formula for the second term in 
Eq. \ref{eq:l_cloud}, which can be conveniently written as:
\begin{equation}
  \label{eq:l_cloud_analytical}
  \frac{f_i^{\mathrm QSO}~\sigma_i}{h}~
  \left(
    \frac{1+z_{\mathrm QSO}}{1+z_{\mathrm k}}~
  \right)^{-\alpha_i^{\mathrm QSO}}~
  \sum_{m = 0}^{\infty} ~
  \frac{(-1)^m}{m!~ (3 m +\alpha_i^{\mathrm QSO}+3)}~
    \left[
      \sum_{l = 1}^{{\mathrm k}-1} ~
      ~ N_i^l ~\sigma_i~
      \left(
        \frac{1+z_{\mathrm k}}{1+z_l} 
      \right)^3
    \right]^m.
\end{equation}

\noindent
In practice, the number of terms to be evaluated  depends on the
number and values of the largest $N_i^l$ lines; we find that a
summation on $m$ up to $\sim 30$ is sufficient for our  purpose.

\subsubsection{Model spectrum}
\label{sec:model_spectrum}

Once the ionization rates are known, we can calculate $\eta$ (Eq. 5) and
 $N_{\mathrm HeII}$.
In the final step, we apply the calculated value of $N_{\mathrm HeII}$
and the line-width ratio $b_{\mathrm HeII}/b_{\mathrm HI}=1$ to obtain
the \ion{He}{2} absorption spectrum (Voigt profiles) of all the lines
in the line list.  After convolution by a line spread function of
FWHM=3.1 pixels (cf. \S\ref{sec:stis_uv_spectra}), we obtain a model
spectrum of the \ion{He}{2} \Lya\,  absorption trough suitable for
comparison with the STIS observations. By adjusting the input
parameters in different simulations, we can measure diagnostic properties of
the UV background and evaluate the need for diffuse gas in the IGM.

\begin{figure}[htp]
  \begin{center}
    \plotfiddle{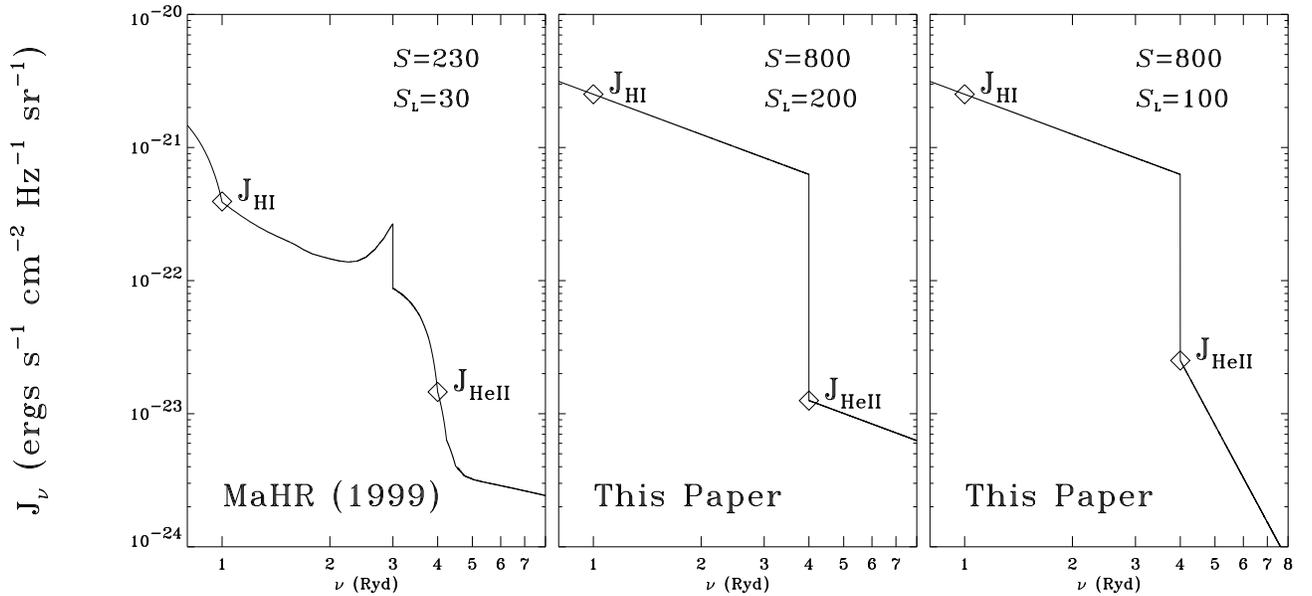}{12cm}{0}{115}{115}{-325}{-130}
    \caption{Comparison between the Madau, Haardt and Rees
  (1999) spectrum of the UV background at $z = 3$ (left panel) and
  broken power laws of index $\beta_{\mathrm HI} = \beta_{\mathrm
    HeII} = 1$ (middle panel) and $\beta_{\mathrm HI} = 1$,
  $\beta_{\mathrm HeII} = 5$ (right panel). For the middle and right
  panels, the values of $J_{\mathrm HeII}$ and $J_{\mathrm HI}$ are
  inferred from the photo-ionization rates necessary to explain the
  proximity effect and the observed opacity in the \ion{He}{2} opacity
  trough, respectively. The Madau et al. (1999) UV background spectrum
  has been multiplied by 1.47 to match our choice of cosmological
  parameters.
\label{fig:comparison_jnu}}
  \end{center}
\end{figure}

To summarize, $\eta(z)$ can be evaluated based on only the observed
values of \NHI, the QSO observed flux and spectral slope, and on some
assumptions about the UV background.  By matching the observed spectrum, 
we can set constraints on the \ion{H}{1}
photo-ionization rate due to the UV background, and on the  softness
parameter,  
\begin{equation}
\label{eq:S}
S \equiv \frac{\Gamma_{\mathrm HI}^{J}}{\Gamma_{\mathrm HeII}^{J}}.
\end{equation}
We prefer to use $S$, which intrinsically takes into account the shape
of the UV background spectrum, instead of another, often used
measure of the shape of the UV background: 
\begin{equation}
\label{eq:S_L}
S_{\mathrm L} \equiv \frac{J_{\mathrm HI}}{J_{\mathrm HeII}},
\end{equation}
where $J_{\mathrm HI}$ and $J_{\mathrm HeII}$ are the
UV background intensities  at the \ion{H}{1} and \ion{He}{2} Lyman
limits, respectively. 
The two quantities are of course proportional,
\begin{equation}
\label{eq:S_propto_SL}
S = r ~ S_{\mathrm L}
\end{equation}
where the  constant of proportionality $r$ takes into account the
exact shape of the UV background spectrum.
If we express the UV background spectrum in terms of a broken power law, then 
from Eq. \ref{eq:photo_ionization_rate_J}, we obtain
\begin{equation}
  \label{eq:r}
  r = 
  \frac{\sigma_{\mathrm HI}}{\sigma_{\mathrm HeII}}  ~
  \frac{3+\beta_{\mathrm HeII}^{\mathrm J}}{3+\beta_{\mathrm HI}^{\mathrm J}}.
\end{equation}
However,   the  latest models   of  the UV background  spectrum depart
significantly    from   a broken    power-law.    For  example, 
Figure~\ref{fig:comparison_jnu} shows the UV background spectrum predicted by
Madau, Haardt \& Rees (MaHR; 1999).  We have numerically evaluated the
integral   given  by  Eq.~\ref{eq:ionization_rate},  using values  for
$J(\nu)$ kindly  provided to us by F.   Haardt.  We obtain $  S \simeq
230$ and $S_{\mathrm L}  \simeq 30$, so  that $r_{\mathrm MaHR} \simeq
8.8$.  \cite{FGS98} (see their Figure  6) present another model of the
UV background   spectrum resulting  from their  source  model Q1  with
spectral  index  $\alpha_{\mathrm  s}  = 1.8$,   stellar contribution,
absorption model A2  and taking cloud re-emission  into account.   The
stellar contribution  is fixed at  1  Ryd to have  an emissivity twice
that of the  quasars and is  considerably softer than the MaHR's model
or other models  without stellar  contribution.  This  Fardal et al.   
model has $S  \simeq 930$, $S_{\mathrm  L} \simeq 140$ and $r_{\mathrm
  FGS} \simeq 6.9$.

We can set constraints on the UV background shape and intensity from the
spectrum of Q~0302--003 for several reasons. 
First, the QSO is bright so that the proximity effect
extends far away from it.
Second, the largest \NHI\, cloud in front of it
has a $\log{N_{\mathrm HI}} = 15.4$, so that even with values of $\eta
\sim 500$, very few clouds have $\log{N_{\mathrm HeII}} > 17$ where
other radiative transfer effects start to be important.  
There is no $z_{\mathrm abs} \simeq z_{\mathrm em}$ system
nearly as strong in \ion{H}{1} absorption as in HE~2347-4342
(\cite{Reimers97}), which leads to continuous absorption shortward of
the \ion{He}{2} edge for that object. The closest major absorber to 
Q~0302--003 at $z=3.2673$ shows only a single \ion{C}{4} component, 
two \ovi\, 
components and a \ion{H}{1} column density of $\log N_{\mathrm HI} =
14.66$.  There are no higher column density systems at $z>3.221$.

Third, the STIS spectrum covers a wide range in redshift, probing both 
the proximity region and regions far from the QSO. 
Without the proximity 
effect, the  observations could not constrain the UV background shape and 
intensity independently. 
Rather, 
the observations 
yield the softness parameter, $S$, which is a function of both  
${J_{\mathrm HI}}/{J_{\mathrm HeII}}$, and the continuum slopes, $\beta_i$.
This ambiguity is also shown in the middle and righthand panels of 
Figure ~\ref{fig:comparison_jnu}. In both panels, the UV background has a softness
$S=800$, but in the righthand panel, ${J_{\mathrm HeII}}$ is two times higher 
which is compensated for by the steeper Lyman continuum slope. With the 
observation of the proximity effect, however, we gain a 
direct measurement of ${\Gamma_{\mathrm HeII}^{J}}$ by itself 
(not the ratio ${\Gamma_{\mathrm HI}^{J}}/{\Gamma_{\mathrm HeII}^{J}}$),
which can then be combined with the observationally derived quantities 
to calculate  ${J_{\mathrm HeII}}$ and ${J_{\mathrm HI}}$.
 

\subsection{Sensitivity of the models to  input parameters}
\label{sec:sensitivity}

Before presenting the results of our modeling, we first summarize the
sensitivity of the model spectra to input assumptions or parameters,
such as the presence/absence of a diffuse gas component 
(\S \ref{sec:diffuse_component}), the UV background \ion{H}{1}
photo-ionization rate $\Gamma_{\mathrm HI}^{\mathrm J}$, and its
spectral shape as described by the softness parameter $S$.  As shown
in Eq.~\ref{eq:S_propto_SL}, $S$ is related to the conventional
softness parameter, $S_{\mathrm L} \equiv J_{\mathrm HI}/J_{\mathrm
  HeII}$, but we work in terms of $S$, because it directly takes into
account the shape of the UV background spectrum.  In \S
\ref{sec:far_QSO}, we will show that far from the QSO, the \ion{He}{2}
\Lya\, spectrum depends only on $S$, but that close to the QSO, it
depends on both $\Gamma_{\mathrm HI}^{\mathrm J}$ and $S$. Finally, we
will also briefly discuss the effect of turbulent vs. Doppler
broadening of the \ion{He}{2} lines in \S \ref{sec:doppler}.

For simplicity, we assume that the slope of the QSO Lyman continuum
evaluated near the \ion{He}{2} Lyman-limit is the same as that of the
\ion{H}{1} Lyman continuum, i.e. $\alpha_{\mathrm HI}^{\rm QSO} =
\alpha_{\mathrm HeII}^{\rm QSO} = 1.9$. We will then discuss the
sensitivity of the results to $\alpha_{\mathrm HeII}^{\rm QSO}$ in \S
\ref{sec:sens_alpha_beta}.

\subsubsection{Presence of a diffuse gas component}
\label{sec:diffuse_component}

Diffuse gas in the form of numerous clouds 
of low column density ($N_{\mathrm HI}<10^{13}~{\mathrm
  cm}^{-2}$) produces a decrease in the transmitted flux in the
\ion{H}{1} spectrum compared to the `observed' line list, which is
only complete down to $N_{\mathrm HI}<10^{13}~{\mathrm cm}^{-2}$.
This diffuse gas is especially important for the \ion{He}{2} \Lya\, 
forest at $z\sim 3$, where the optical depth is larger than unity for
large values of the softness parameter.

\subsubsection{Dependence of $\eta$ on $\Gamma_{\mathrm HI}^{\mathrm
    J}$ and $S$} 
\label{sec:far_QSO}

Far from the QSO, where the second term of Eq.
\ref{eq:l_cloud_analytical} is negligible compared to the first one,
the value of the ratio of column densities $\eta = N_{\mathrm
  HeII}/N_{\mathrm HI}$ is only proportional to $S$, through Eq.
\ref{eq:eta_44}.
This region of the spectrum can be used to constrain
the spectral shape of the UV background.

Close to the QSO, the second term in Eq. \ref{eq:l_cloud} becomes
important.  For reference, the \ion{H}{1} photo-ionization rates due
to the UV background and the QSO are equal at $z = 3.268$ if
$\log{J_{\mathrm HI}} = -21$, $\beta_{\mathrm HI}^{\rm J} =1$, and for
the QSO \ion{H}{1} Lyman limit luminosity shown in
Fig.~\ref{fig:hubenyfig1}.  The same will be true for the \ion{He}{2}
photo-ionization rates if $S_{\mathrm L} = 4^{\alpha^{\mathrm QSO}} = 13.9$ (and
if $\beta_{\mathrm HI}^{\rm J} = \beta_{\mathrm HeII}^{\rm J} = 1$).
Decreasing $\Gamma_{\mathrm HI}^{\mathrm J}$ while holding constant
the softness parameter, $S$, extends the redshift range over which the
QSO has a significant effect on both \ion{H}{1} and \ion{He}{2}.
Increasing $S$ while holding $\Gamma_{\mathrm HI}^{\mathrm J}$
constant has the same effect but only for \ion{He}{2}, since this
manipulation actually corresponds to decreasing $\Gamma_{\mathrm
  HeII}^{\mathrm J}$.  Since decreasing $\Gamma_{\mathrm HI}^{\mathrm
  J}$ and increasing $S$ have the same effect on the \ion{He}{2}
spectrum, the proximity effect can constrain the softness parameter if
$\Gamma_{\mathrm HI}^{\mathrm J}$ is known. Similarly, limits on
$\Gamma_{\mathrm HI}^{\mathrm J}$ can be derived from $S$ which can be
estimated from the run of $\eta$ far from the QSO.

\subsubsection{Doppler broadening}
\label{sec:doppler}

As mentioned in \S \ref{sec:n_heii}, we only consider turbulent line 
broadening ($b_{\mathrm HeII}/b_{\mathrm HI}=1$) to
calculate the \ion{He}{2} absorption spectrum.  If instead, thermal
motions were the dominant broadening mechanism ($b_{\mathrm
  HeII}/b_{\mathrm HI}=0.5$), then the \ion{He}{2} opacity would be
lower, and the UV background would have to be even softer in order to
reproduce the observations.

\subsubsection{Sensitivity of results to the QSO flux distribution}
\label{sec:sens_alpha_beta}

The value of $\alpha_{\mathrm HI}^{\rm QSO} = 1.9$ is secure, but
stating that $\alpha_{\mathrm HeII}^{\rm QSO} = 1.9$ results from an
extrapolation.  To our knowledge, the \ion{He}{2} Lyman continuum
region has never been observed, even in low-redshift quasars.  As
noted in \S \ref{sec:qso_spectrum}, a softer QSO spectrum is
theoretically possible but improbable since high-ionization absorption
lines are detected in quasar spectra.  If the spectrum were as soft as
given by $\alpha_{\mathrm HeII}^{\rm QSO} = 11$, then the QSO
contribution to the photo-ionization rate of a cloud would be three
times lower than for $\alpha_{\mathrm HeII}^{\rm QSO} = 1.9$, and the
proximity zone would be smaller.
In order to match the observations, $\Gamma_{\mathrm HeII}^{\mathrm
  J}$ would have to be smaller, and the softness parameter $S$
increased by a factor of 3.

\subsection{Results} 
\label{sec:results}

We now present the results of modeling \ion{He}{2} Gunn-Peterson
absorption in the spectrum of Q~0302--003. 
Comparing the modeled spectra with the observations allows us: (1) to
confirm the absence of a diffuse gas component close to the quasar (\S
\ref{sec:nodiffusecpt}); (2) to constrain the values of $ J_{\mathrm
  HI}$ and $S$ from the proximity effect (\S
\ref{sec:jsproxestimates}); (3) to show that diffuse gas component
must be present far from the quasar, and to set limits on $\eta$ and
$S$ (\S \ref{sec:etafarfromqso});
(4) to show evidence for a soft UV background (\S
\ref{sec:evidence_soft_UVB}), which allows the UV background flux at
the \ion{H}{1} Lyman limit $ J_{\mathrm HI}$ to be determined (\S
\ref{sec:determination_of_JHI}). 

In practice, the calculations were performed using $\beta_{\mathrm
  HI}^{\mathrm J} = \beta_{\mathrm HeII}^{\mathrm J} = 1$ for the
slopes of the UV background just blueward of the Lyman limits, and  we
present the models in terms of $ J_{\mathrm HI}$ and $S$ for easy
comparison with other studies.  Since the the fundamental variables
are the photo-ionization rates, an easy conversion to  $ J_i$
for other values of $\beta_i^{\mathrm J}$ can be done using Eq.
\ref{eq:gamma_J}.

\subsubsection{No diffuse gas component close to Q~0302--003 }
\label{sec:nodiffusecpt}

\begin{figure}[htbp]
  \begin{center}
    \plotfiddle{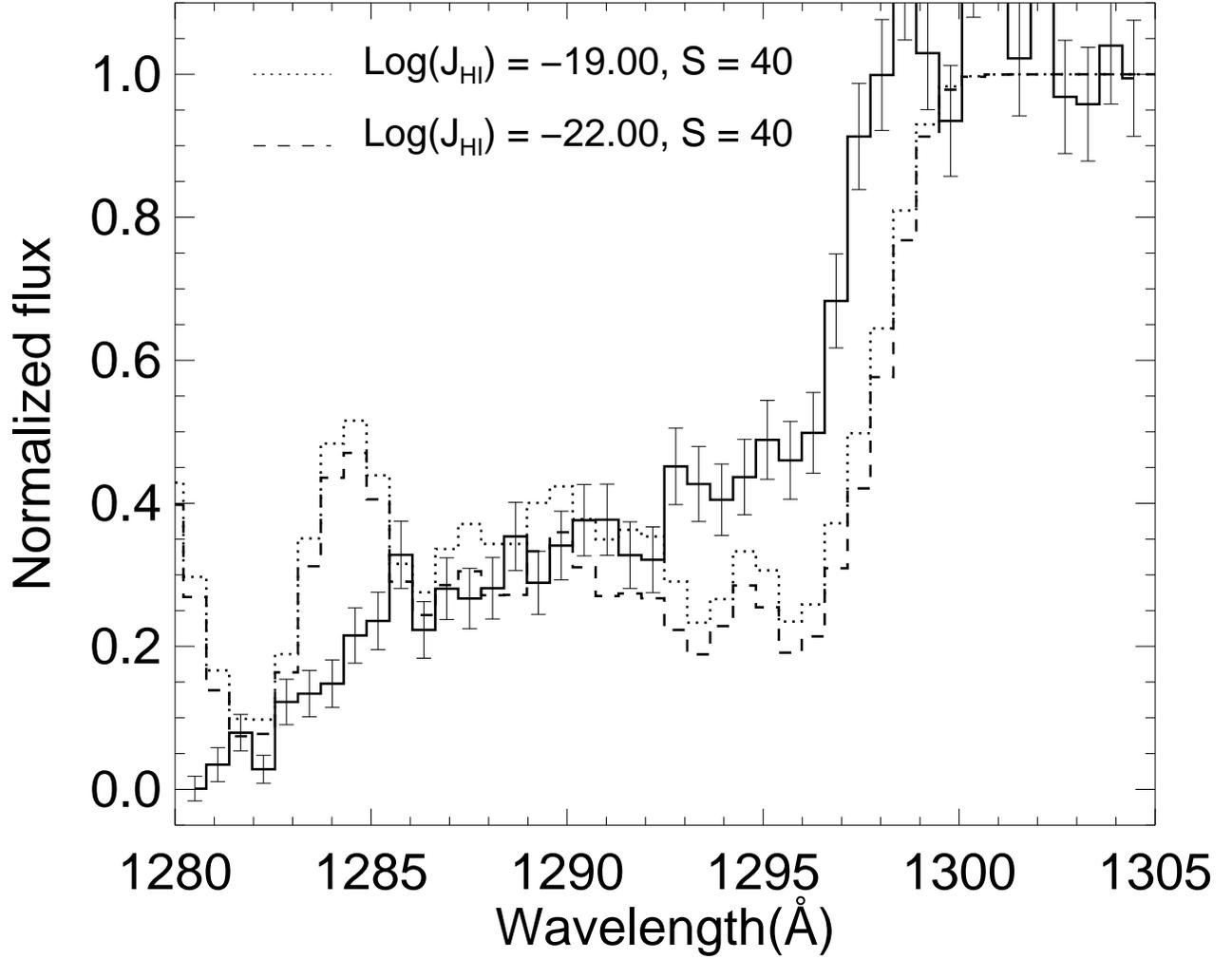}{12cm}{90}{85}{85}{300}{-40}
    \caption{Attempt to reproduce the observed spectrum
  (solid) using the combined line list (dotted: $J_{\mathrm HI} = 1
  \times 10^{-19}~{\mathrm erg}~{\mathrm s}^{-1}~{\mathrm
    cm}^{-1}~{\mathrm Hz}^{-1}~{\mathrm sr}^{-1}$, dashed: $J_{\mathrm
    HI} = 1 \times 10^{-22}~{\mathrm erg}~{\mathrm s}^{-1}~{\mathrm
    cm}^{-1}~{\mathrm Hz}^{-1}~{\mathrm sr}^{-1}$). The `best' match
  is obtained with $S = 40$, as shown here.
\label{fig:diffuse}}
  \end{center}
\end{figure}

We explored a wide range in parameter space ($J_{\mathrm HI}$, $S$) 
and found no way to reproduce the observed
quasar spectrum in the proximity zone (1280 -- 1300~\AA ) if we use
the \textit{combined line list}, i.e. if we allow the presence of a
diffuse gas component.  Even for a UV background spectrum that is
harder than the QSO's (cf. Figure~\ref{fig:diffuse}), the
\mbox{1297~\AA\, } edge is too depressed, and the region between
\mbox{1285~\AA\, } and \mbox{1297~\AA\, } is too flat compared to the
observed spectrum.  On the other hand, we can achieve a satisfactory
fit if we use only the observed line list. We therefore agree with
Hogan et al. (1997) that there is no diffuse component close to the
QSO.

\subsubsection{Constraint on $J_{\mathrm HI}$ 
and $S$ from the  proximity effect}
\label{sec:jsproxestimates}

Here, we focus on reproducing the STIS spectrum in the proximity zone
of the QSO, i.e. $\lambda=1282 - 1301$~\AA.  A large number of
observed pixel fluxes can be reproduced by a wide range of values of
$J_{\mathrm HI}$ and $S$ and thus are not useful for
their precise determination.  There are, however, two regions of $\sim
5$-pixel width, centered at $\sim$1285~\AA\, and $\sim$1290~\AA , that
are quite sensitive to $J_{\mathrm HI}$ and $S$.
Indeed, in the absence of a diffuse component, the predicted values of
\NHEII\,  for lines in these regions imply a $\tau_{\mathrm HeII} \sim
1$ in the cores of the lines for the values of $\Gamma_{\mathrm
  HI}^{\mathrm J}$ and $S$ of interest.  Hence, these regions provide
the best way to estimate $J_{\mathrm HI}$ and $S$.
The opacity in the $1285$ \AA\, region is somewhat larger than in the
$1290$ \AA\, region, so that satisfactory fits are only obtained with a
relatively small set of parameters.  However, it is possible that the
1285 \AA\, region is affected by an ionizing source that could be also
responsible for the Dobrzycki-Bechtold void seen in \ion{H}{1} (see
below). If so, the $1290$ \AA\, region provides the stronger constraint
on the UV background spectrum.  Finally, we note that there are other
spectral intervals within the proximity zone that have line-core
opacities close to 1, but at the STIS resolution, these intervals are
contaminated by strong lines nearby. Spectra at higher resolution are
needed to better define these spectral intervals with $\tau \sim 1$
and thus allow a better determination of $\Gamma_{\mathrm HI}^{\mathrm
  J}$ and $S$.

\begin{figure}[htbp]
  \begin{center}
    \plotfiddle{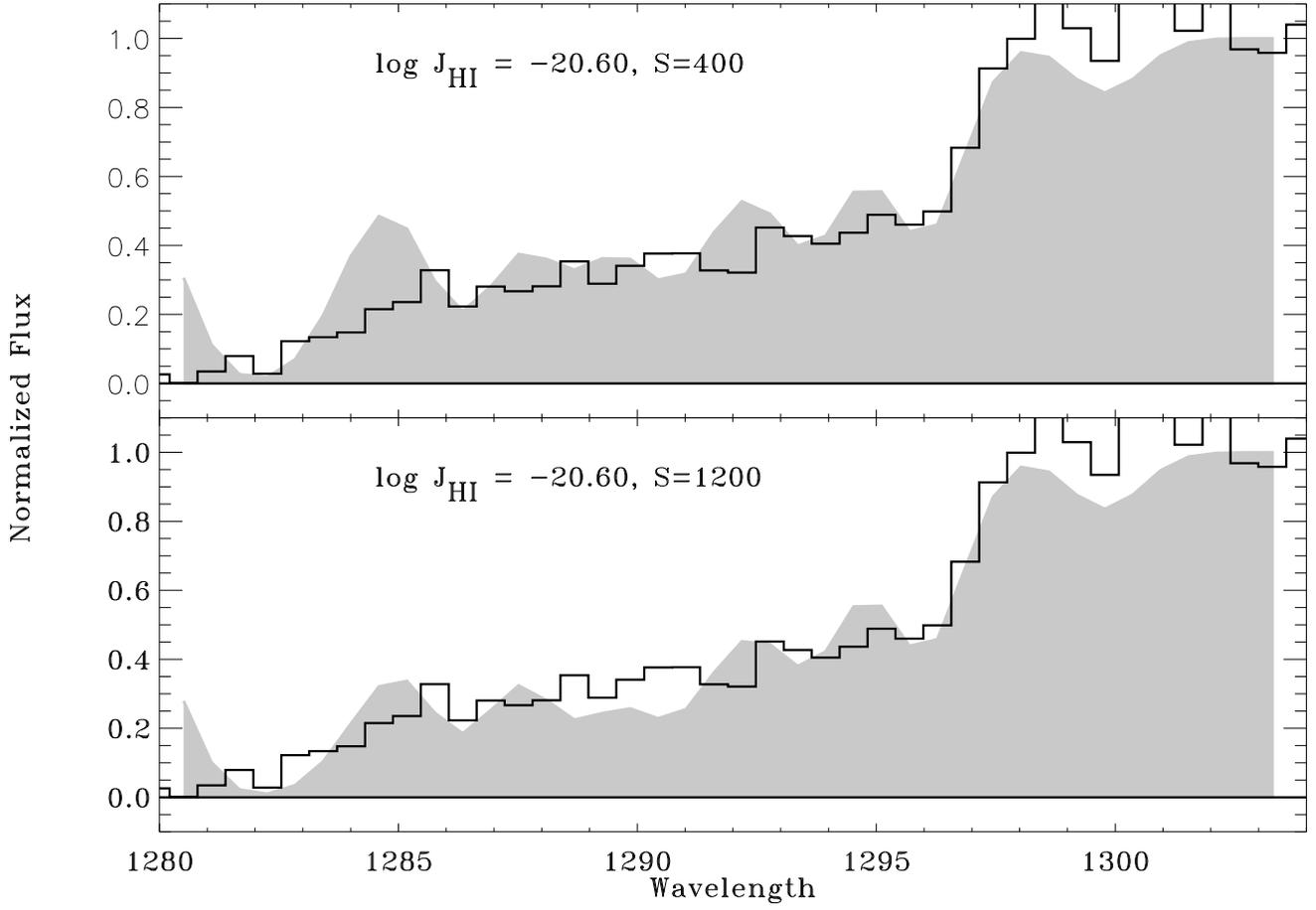}{12cm}{0}{100}{100}{-310}{-310}
    \caption{Models of the proximity effect region for two
  values of the softness parameter $S = \Gamma_{\mathrm HI}^{\mathrm
    J}/\Gamma_{\mathrm HeII}^{\mathrm J}$ (in gray). The solid line
  represents the normalized STIS spectrum.
\label{fig:proximity}}
  \end{center}
\end{figure}

Figure~\ref{fig:proximity} compares two model spectra with the
normalized observed spectrum. The two models, with their different
values of $S$, bracket a best model of the proximity effect.  If
$J_{\mathrm HI}$ were fixed at a lower value, e.g.  $1 \times
10^{-21}~{\mathrm erg}~{\mathrm s}^{-1}~{\mathrm cm}^{-2}~{\mathrm
  Hz}^{-1}~{\mathrm sr}^{-1}$, then a very large value of $S \ga 4000$
would be required.  On the other hand, the background radiation field
could be as high as $J_{\mathrm HI} = 1\times10^{-18}~{\mathrm
  erg}~{\mathrm s}^{-1}~{\mathrm cm}^{-2}~{\mathrm Hz}^{-1}~{\mathrm
  sr}^{-1}$ if $400 < S < 800$.

\subsubsection{Determination of $S$ far from the quasar}
\label{sec:etafarfromqso}

We now estimate the softness parameter $S$ at lower redshifts. It
quickly appears that a diffuse component is needed to reproduce the
high \ion{He}{2} opacity in the D--B region. We thus created a line
list that contains only the observed lines at $z > 3.220$ (i.e. in the
proximity zone) and all the lines from the `combined' line list at $z
< 3.220$. This boundary redshift corresponds to the redshift of the
strongest \ion{H}{1} absorber in the spectrum of Q~0302-003.  As this
absorber is relatively close to the quasar, it efficiently blocks
\ion{He}{2} ionizing flux from the quasar.

\begin{figure}[htbp]
  \begin{center}
    \plotfiddle{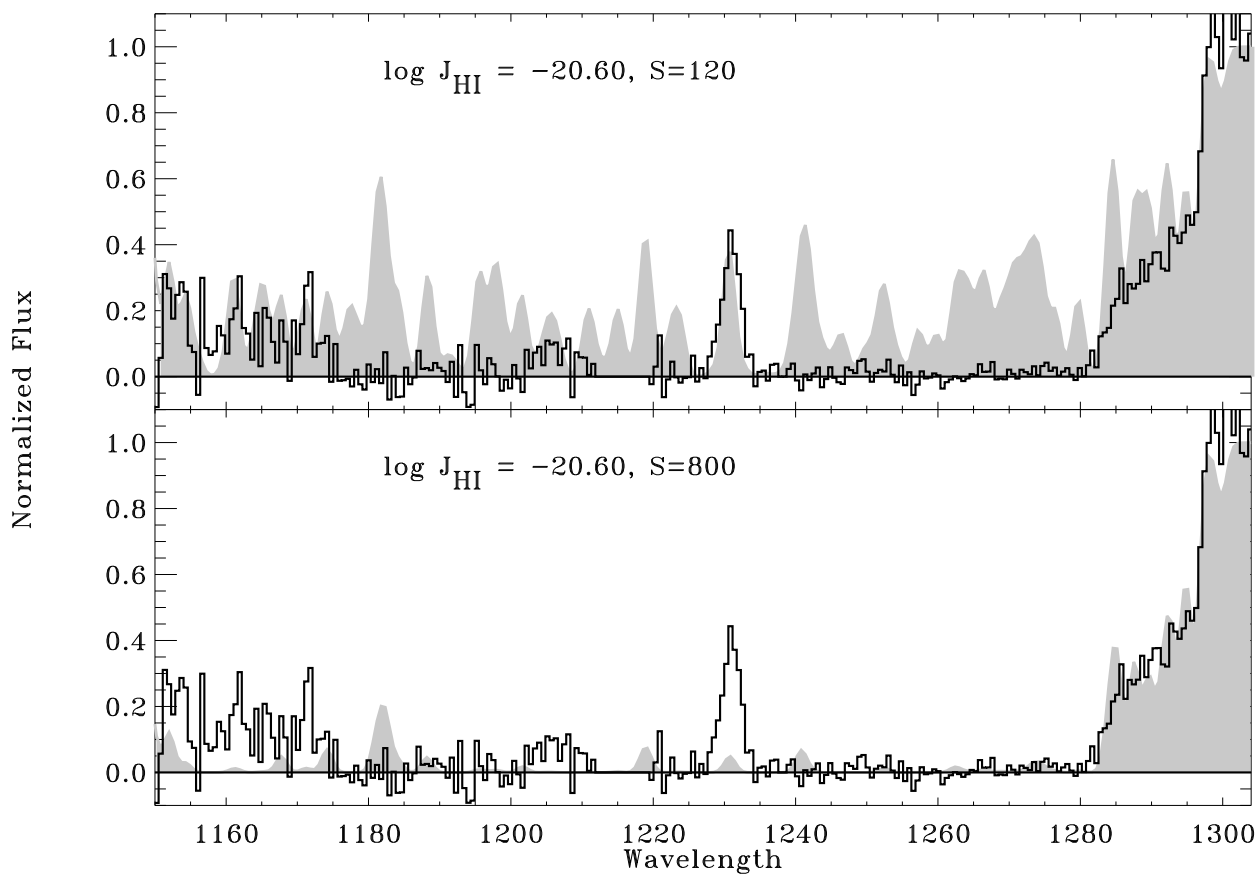}{12cm}{0}{95}{95}{-300}{-310}
    \caption{Model \textit{vs} observed spectrum. The
  thick solid line represents the observed spectrum. The gray regions
  shows the model spectra:  {\it above:} with $S = 120$ that best matches the
  low opacity region at $z=3.05$ and produces an acceptable fit in the
  regions $1150~{\mathrm \AA } < \lambda < 1175~{\mathrm \AA }$; {\it below:}
  with $S= 800$ (and
  $\log{J_{\mathrm HI}} = -20.6$) giving a good fit over most of
  the observed spectral range.
\label{fig:model}}
  \end{center}
\end{figure}

Figure~\ref{fig:model} compares the observed spectrum with model
spectra computed for two different values of the softness parameter
$S$.  It shows that a hard UV background with $S\sim 120$ produces a
good match to the \ion{He}{2} opacity gap at $\lambda = 1230$~\AA\, ($z
= 3.05$), and it provides a reasonable fit to the low-redshift regions
of the spectrum, $1150~{\mathrm \AA } < \lambda < 1175~{\mathrm \AA
  }$. The region, $1205~{\mathrm \AA } < \lambda < 1210~{\mathrm \AA
  }$, needs a somewhat softer background with $S\sim 300$. However, the
UV background must be quite soft, i.e.  $S= 800-1000$, to explain the
observed opacity over most of the STIS spectrum including the
Dobrzycki-Bechtold `void' (the D--B region).  An even softer UV
background produces too large an opacity: if $S \sim 1400$, the model
produces only 1/6 of the flux observed in Region A.

We stress that the method here automatically takes into account
changes in the gas density traced by the Ly$\alpha$ clouds, as long as
most of the hydrogen is ionized and helium is doubly ionized.
Consequently, the features that our model spectra cannot reproduce
with a unique value of the softness parameter are likely due to local
changes in the ionizing flux.  For example, the fact that the strong
\ion{He}{2} opacity in the Dobrzycki-Bechtold `void' can be reproduced
by a diffuse component suggests a nearby source that is able to
completely ionize the hydrogen but unable to doubly ionize the helium.
Such a source must have a much softer spectrum than a typical QSO.
Other examples include the low \ion{He}{2} opacity at $1230$~\AA\, ($z
= 3.05$) which corresponds to a local decrease in the $J_{\mathrm
  HI}/J_{\mathrm HeII}$ ratio, and suggests an AGN close by, or some
overlapping Str{\"o}mgren spheres of more distant sources. The
marginal decrease in the \ion{He}{2} opacity at 1207~\AA\, may be
similarly explained. In addition, some regions of low \ion{He}{2} opacity
like the one at $\sim 1182$~\AA\, are expected in our model; we discuss
them in \S\ref{sub:underdensevoids}.  Finally, the extent of the region
$1150~{\mathrm \AA } < \lambda < 1175~{\mathrm \AA }$ and the
observation of HS~1700+6416 (Davidsen et al. 1996) suggest that the
UV background presents a harder spectrum at $z < 3$ at which the
Universe is effectively transparent.

\begin{figure}[htbp]
  \begin{center}
    \plotfiddle{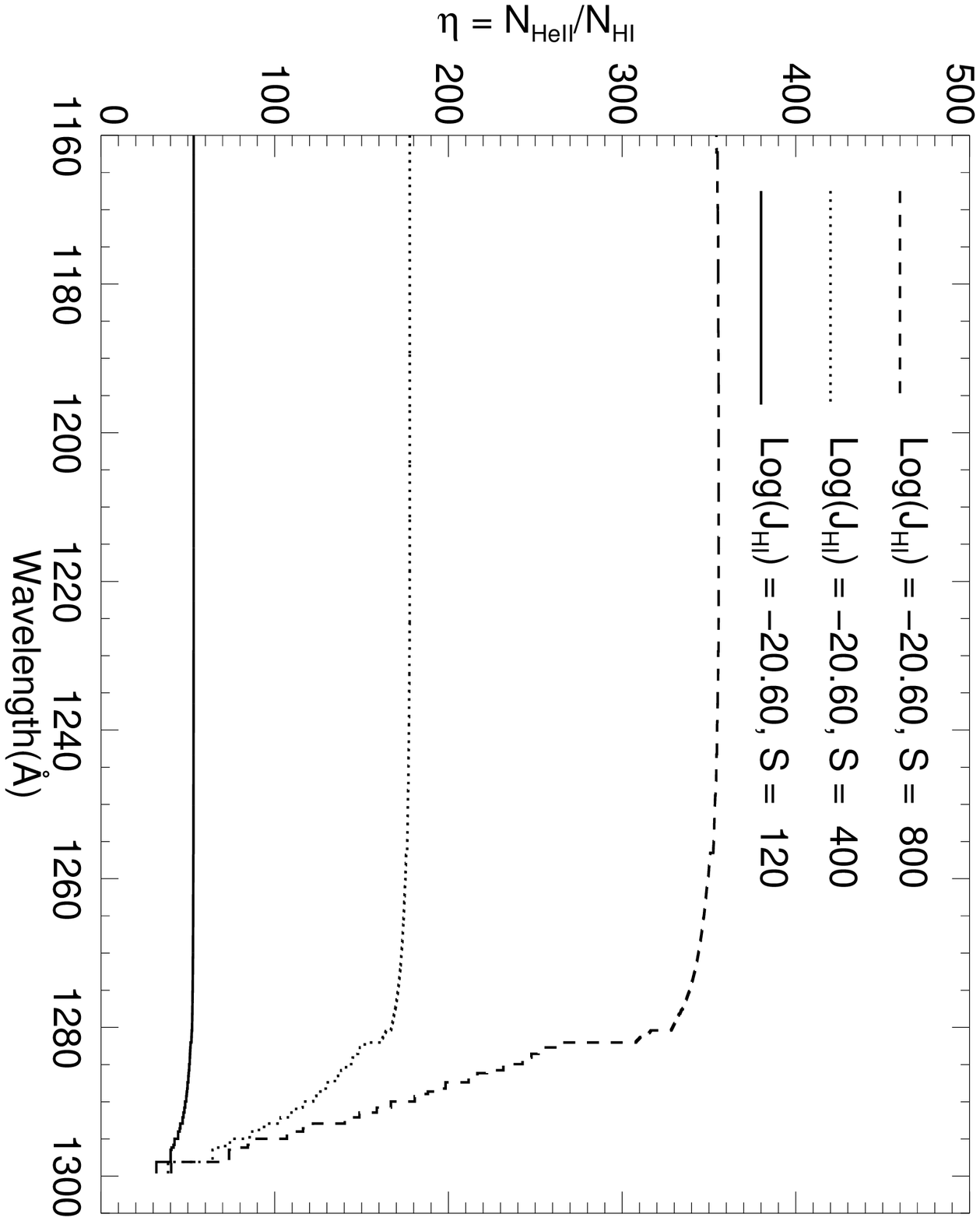}{13cm}{90}{85}{85}{315}{-35}
    \caption{Behaviour of $\eta$ as derived by our model
  using the 'observed' line list in the proximity region and the
  'combined' line list far from the quasar.  Three different values of
  $S$ are presented for $\log{J_{\mathrm HI}} = -20.60$, although only
  the $ S = 800$ case models the data well. Note that lower values for
  $J_{\mathrm HI}$ are allowed, but require even larger values for
  $S$, so that $\eta \simgt 200 $ far from the quasar in all the cases
  which reproduce the data acceptably.
\label{fig:eta_prox}}
  \end{center}
\end{figure}

Figure~\ref{fig:eta_prox} shows how the ratio of column densities
$\eta = N_{\mathrm HeII}/N_{\mathrm HI}$ varies as a function of
wavelength (redshift). All the curves assume the same intensity of the
UV background, $\log{J_{\mathrm HI}} = -20.60$, and use the same line
list (observed line list in the proximity zone, combined line list
further away from the QSO). They differ only in the softness of the UV
background. In all cases, $\eta$ increases gradually with distance
from the QSO. A similar figure for larger $J_{\mathrm HI}$ would
display flatter curves.  In all the cases which reproduce the data
well, $\eta \simgt 350 $ far from the quasar, as shown by the top
curve.

\subsubsection{Evidence for a soft UV background}
\label{sec:evidence_soft_UVB}
From  the  previous  sections, and   as  also  inferred from  previous
observations of the \ion{He}{2}  Gunn-Peterson effect, we  find little
support  for  a   `hard'  ($S_{\mathrm  L}\sim30$)   UV background  as
predicted by Haardt \& Madau (1996) and  Madau, Haardt \& Rees (1999). 
Instead, in addition    to the direct measurement of    $\tau_{\mathrm
  HeII}$ described    in section~\ref{sec:heiigptrough},  our modeling
shows  evidence for a soft background  both from  the proximity effect
and  from the  redshift domain away  from  the quasar ionizing region. 
The value of the  softness parameter $S \simeq  800$  is close to  the
value $\simeq 930$ derived for  the Fardal et  al. (1998) model (their
Figure 6, source  model Q1 with  spectral index $\alpha_{\mathrm  s} =
1.8$, stellar contribution fixed at 1 Ryd  to have an emissivity twice
that of the quasars, absorption  model A2 and taking cloud re-emission
into account), which suggests that  a significant stellar contribution
is required.

In order to compare the ratio $S_{\mathrm L} = J_{\mathrm
  HI}/J_{\mathrm HeII}$ of \ion{H}{1} to \ion{He}{2} UV background
intensities with other values found in the literature, we have to
assume a shape for the UV background spectrum. For a broken power-law
model with $\beta_{\mathrm HI} = \beta_{\mathrm HeII} = 1$, the
proportionality constant $r = 4$ (Eq. \ref{eq:r}) so that $S_{\mathrm
  L} = 200$. Instead, if we use the shape of the spectrum obtained by
Fardal et al. (1998), $r = r_{\mathrm FGS} \simeq 6.9$ (\S
\ref{sec:model_spectrum}) and $S_{\mathrm L} \sim 120$.

\subsubsection{A determination of $J_{\mathrm HI}$}
\label{sec:determination_of_JHI}
Since a softness parameter $S \simeq 800$ is required by our models
over most of the range of the STIS spectrum, the constraint imposed by
the proximity effect sets $\log{J_{\mathrm HI}} = -20.6$
for $\beta_{\mathrm HI}= 1$ (cf. Fig. \ref{fig:comparison_jnu}).

Taking into account the differences in the assumed cosmological
parameters and UV background spectral slopes between the ones we
assume and those chosen in the following studies, this value of the UV
background is smaller than the 3$\sigma$ upper limit derived by a
recent attempt to detect \lya\ emission in Lyman-limit
absorption systems (\cite{Bunker98}), but is a factor of 2 larger than
the one obtained by \cite{Scott98} and a factor of $\sim 3$ larger
than the one derived by \cite{Giallongo96} from the study of the
Ly$\alpha$ forest.

\section{THE OPACITY GAP AT {\boldmath $z=3.05$}}
\label{sec:voidz305}

We now return to the windows of transmitted flux seen in all three sightlines 
probed by STIS (cf. Figure \ref{fig:srhfig5}). These windows give insight into
the reionization history of the IGM and how the reionization of He occurred.
Indeed, their observed properties provide a direct point of comparison
with the growing number of theoretical studies that predict the
\ion{He}{2} opacity with respect to the UV background flux, IGM
density and other physical aspects of the early universe
(\cite{Croft97}; \cite{FGS98}; \cite{Zheng95}; \cite{Zhang98};
MHR and references therein).

The results of our models described in \S 5 imply that the opacity gap
at $z=3.05$ is most likely produced by a local source able to doubly
ionize helium, or the result of a change in the $J_{\mathrm
  HI}/J_{\mathrm HeII}$ ratio (due, for example, to additional
ionizing sources or to increased transparency of the medium to
\ion{He}{2} Lyman continuum radiation). In this section, we analyze
this opacity gap in more detail. We first describe in \S 6.1 the
observed properties of these regions of enhanced transmitted flux seen
in the Q~0302-003 spectrum.  We then examine three proposed
explanations of \ion{He}{2} opacity gaps.  In
\S\ref{sub:underdensevoids}, we consider the possibility that 
low opacities are
the signatures of voids, or underdense regions ($\rho/\bar{\rho} \sim
10^{-1} $).  
For our second interpretation, we explore in
\S\ref{sec:voidadditionalionizer}  the hypothesis that
opacity gaps are associated with regions in which helium has been
doubly ionized by nearby discrete sources.  Here, we have used the
estimated UV background from \S\ref{sec:modeling} to test whether an
AGN is required near the $z=3.05$ gap.
Finally, in \S 6.4, we briefly examine the possibility that
opacity gaps are regions that have been collisionally ionized by
shock--heated gas.

\begin{deluxetable}{lcccccc}
\tablecolumns{7}
\tablecaption{HeII Absorption Gaps}
\tablehead{
\colhead{ Sightline        } &  
\colhead{ $z_{\mathrm gap}$        } &  
\colhead{$\Delta D$} &
\multicolumn{3}{c}{Linear extent} & 
\colhead{Instrument} \\  \cline{4-6}
\colhead{} &
\colhead{} & 
\colhead{(comoving Mpc)} & 
\colhead{(\AA )} & 
\colhead{(\kms )} & 
\colhead{(comoving Mpc)} & 
\colhead{} \\
}
\startdata
Q~0302--003   & 3.052 &     202.5 & 5.8 &     1420 &     17.2 & STIS \\
PKS~1935--692 & 3.100 & \phn 69.8 & 4.7 &     1130 &     13.6 & STIS \\
HE~2347--4342 & 2.865 & \phn 19.0 & 2.3 & \phn 590 & \phn 7.2 & GHRS \\
HE~2347--4342 & 2.814 & \phn 67.9 & 4.0 &     1043 &     12.8 & GHRS \\
\enddata
\label{tab:bigvoids}
\end{deluxetable}

\begin{deluxetable}{lcccl}

\tablecolumns{4}
 
\tablecaption{\ion{H}{1}, metal absorbers near the $z=3.0526$ \ion{He}{2} opacity gap}
\tablehead{
\colhead{ion}&
\colhead{$z$}&
\colhead{$b$ (\kms) }&
\colhead{log $N_{\mathrm ion}$ (cm$^{-2}$)\tablenotemark{a}}\\
}

\startdata
\ion{C}{4}  & $3.046980\pm 0.000018$ & \phantom{1}$ 6.2\pm 2.2$ & $12.22\pm 0.11$\nl
\ion{C}{4}  & $3.047275\pm 0.000015$ &            $11.4\pm 2.0$ & $12.62\pm 0.05$ \nl
\ion{H}{1}  & $3.045994\pm 0.000039$ &            $28.1\pm 2.6$ & $14.50\pm 0.09$ \nl
\ion{H}{1}  & $3.047073\pm 0.000036$ &            $63.8\pm 2.0$ & $15.32\pm 0.02$ \nl
\ion{H}{1}  & $3.048479\pm 0.000039$ &            $35.6\pm 2.0$ & $14.29\pm 0.06$
\enddata
\tablenotetext{a}{Upper limits ($4\sigma$, assuming $z=3.0472$) are:
\ion{C}{3}~$\lambda 977$   $\leq 11.45$, 
\ion{C}{2}~$\lambda 1335$  $\leq 11.93$, 
\ion{Si}{4}~$\lambda 1393$ $\leq 11.35$,
\ion{Si}{3}~$\lambda 1206$ $\leq 11.05$,
\ion{Si}{2}~$\lambda 1526$ $\leq 12.06$,
\ion{N}{5}~$\lambda 1238$  $\leq 11.93$,
\ion{O}{6}~$\lambda 1031$  $\leq 12.26$.
}

\label{tab-civsystems0302}
\end{deluxetable}

\renewcommand{\arraystretch}{.5}
\begin{deluxetable}{lcccl}
\tablecolumns{5}
\tablecaption{High-Redshift Emission Line Galaxy Surveys}
\tablehead{
\colhead{redshift}&
\colhead{density}&
\colhead{$3\sigma$~flux }&
\colhead{No.}&
\colhead{Reference} \\
\colhead{}&
\colhead{($10^{-4}$ Mpc$^{-3}$)\tablenotemark{a}} &
\colhead{($10^{-16}$ erg cm$^{-2}$ s$^{-1}$)\tablenotemark{b}}&
\colhead{found} &
\colhead{} \\
\colhead{}&
\colhead{}&
\colhead{}&
\colhead{}&
\colhead{} \\
}
\startdata
2.3--2.4, 0.89              &\phm{$<11$}6\phm{.2}&  \phn4.8\phn    & 18     & Mannucci et al. 1998 \nl
2.3--2.5  &\phm{$<1$}41\phm{.2}&  \phn1.0\phn     & \phn5  & Teplitz et al. 1998  \nl
3.4                        &\phm{$<1$}17\phm{.2}&  \phn0.09        & 12     & Cowie \& Hu 1998  \\  
\enddata
\tablenotetext{a}{Comoving space density and star formation rate were
  calculated for our cosmological parameters.}
\tablenotetext{b}{Only objects with line fluxes greater than $1\times
  10^{-16}$~erg~cm$^{-2}$~s$^{-1}$ were included in the tally.}
\label{tab-galsurveys}
\end{deluxetable}
\renewcommand{\arraystretch}{1.0}

\subsection{Observed properties of \ion{He}{2} opacity gaps}
\label{sub:gap_properties}

Table \ref{tab:bigvoids} gives some of the measured characteristics of
the observed opacity gaps in all three sightlines probed by HST. 
Successive columns list the redshift
of the opacity gap and its comoving distance from the QSO
($\Delta D$), the size of the gap (in~\AA , \kms , and comoving
Mpc), and finally, the HST instrument used to obtain the observation.
For consistency, the redshift and widths of the two gaps in the
spectrum of HE~2347-4342 were measured from GHRS data rebinned to the
(lower) resolution of STIS.  The gap size is defined as the interval
of continuous pixels with flux $>1 \sigma$, except for PKS~1935--692,
where the flux redward of the gap is not zero.  In that case, the
boundary of the 
gap was taken as the wavelength at which the flux equals the mean flux
over 1248--1264~\AA.

Since all three lines of sight observed by HST show opacity gaps, we
assume that the line of sight to Q~0302--003 is typical. We can
therefore calculate the filling factor of the region responsible for
the $z=3.05$ opacity gap as the ratio of the path-length through the
gap, $\Delta \ell \simeq 17.2$ comoving Mpc, with respect to the total
path-length probed by the STIS spectrum, $\Delta L$. 
We obtain $\Delta L = 395$ comoving Mpc if we integrate over the
observed Gunn-Peterson trough ($2.7609<z<2.9879$ and
$3.0142<z<3.2183$; that is, avoiding the proximity-effect region and
the region contaminated by geo-coronal \Lya ), or $\Delta L = 180$
comoving Mpc, if we only integrate down to $z=3.0142$, 
just longward of geo-coronal \Lya .
The resulting filling factors are 0.04 and 0.10 respectively.  If we
include all three sightlines, which probe a total of 
$\Delta L \simeq 560$ comoving
Mpc, then we obtain a filling factor of 0.09.  If we assume that each
region of enhanced transmission is a sphere whose diameter is given by
the average size of the opacity gaps, 
$d = 13 $ comoving Mpc, then the volume probed by the spectra is 
$\pi ~ (d/2)^2 ~\Delta L \simeq 7.4 \times 10^{4}$ comoving
Mpc$^3$, and the number density of opacity gaps is $n_{\mathrm gap} \sim
5\times 10^{-5}$ Mpc$^{-3}$.

\subsection{\ion{He}{2} opacity gaps as low-density regions}
\label{sub:underdensevoids}

Cosmological hydro-dynamical simulations predict growing density
fluctuations with time, and MHR interpret opacity gaps in the \ion{He}{2} 
Gunn-Peterson absorption trough as the spectral signatures of  
underdense regions.  Below, we test their interpretation by comparing the 
predicted properties of low-density regions to the observed widths and 
amplitudes of opacity gaps.  

\textit{Sizes of underdense regions.} MHR modeled the effects of a
clumpy IGM on the reionization history of helium. They concluded that
there were sufficient photons to ionize most of the helium by $z=3$
unless very luminous QSOs were the only sources of \ion{He}{2}
ionizing photons.  Based on the \ion{He}{2} \Lya\ opacities we
measure, they estimate that at $z=3$, helium is doubly ionized up to
an overdensity of $\rho/\bar{\rho} \approx 12$, with a volume fraction
$F_V\approx 0.9943$ of helium doubly ionized, and a mean free path of
4-Ryd photons of $\sim 1500$ \kms .  Fluctuations in the density and
ionizing background should produce opacity fluctuations with a similar
scalelength.  The predicted mean free path is consistent with the
largest of the \ion{He}{2} opacity gaps (cf.  Table~3).

\textit{Transmission of underdense regions.} 
In \S \ref{sec:etafarfromqso}, we noted that gaps in the
\ion{He}{2} opacity appear in our simulations due to random
fluctuations in the diffuse IGM (i.e. $\log N_{\mathrm{HI}} \le 13$).
An example of such an opacity gap can be seen
at $\lambda \sim 1182$~\AA\, ($z = 2.890$) in Fig.~\ref{fig:model}.  Could
similar fluctuations produce the prominent $z=3.05$
opacity gap, which has optical depths as low as
$\tau = 0.5$?  To answer this question, we obtained 100
realizations of the simulations described in \S \ref{sec:modeling}
using $S = 800 $ and $\log J_{\mathrm{HI}} = -20.6$.  As mentioned in
\S \ref{sec:jsproxestimates}, this value of $S$ 
can
account for the large
opacities in the Gunn-Peterson trough; lower values of $J_{\mathrm{HI}}$
lead to larger $S$ values.
We then measured the distribution of opacity gaps produced by our
simulations. We found that none of our 100 simulations produced a
gap having an 
optical depth smaller than
$\tau = 1.3$ over the whole spectral range covered by our STIS
spectrum. Instead, the simulations produced an average of
1.2 opacity gaps with optical depths $\tau \le 2.0$.
Most occur at low redshift  where the number density \ion{H}{1} \lya\ lines 
is low, or less frequently, near the redshift of the QSO where
excess ionization by the QSO is not negligible. In contrast,
\textit{the observed opacity gap at $z = 3.05$ 
is located in a region unlikely to show an  opacity gap caused by 
  fluctuations in the diffuse IGM}. In fact, only one of our simulations
produced a gap having an optical depth smaller than $\tau = 1.8$ in the
redshift range $3.0 < z < 3.1$.

In summary, the predicted sizes of underdense regions are compatible
with the observed widths of the opacity gaps. However, our simulations
indicate that the underdense regions cannot produce the distribution
and amplitude of the observed transmission.  This failure constitutes
a strong argument against the hypothesis that the $z = 3.05$ gap
arises as a fluctuation in the diffuse gas component of the IGM.

\subsection{\ion{He}{2} opacity gaps produced by discrete ionizing sources}
\label{sec:voidadditionalionizer}

\begin{figure}[htp]
  \begin{center}
    \plotfiddle{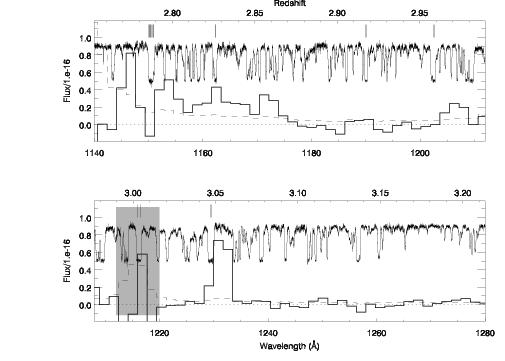}{13cm}{0}{110}{110}{-370}{-390}
    \caption{Location of CIV absorbers toward Q~0302--003.
The STIS spectrum of Q~0302--003, binned by 3 pixels for clarity, 
is shown by the thick solid line.  The 
optical HIRES spectrum (thin solid line) has been divided by a continuum
fit to remove the \Lya\,  emission line, scaled by $2.5\times 10^{-16}$
to match the STIS spectrum redward of 1300~\AA , and offset by
$1\times 10^{-16}$ for clarity. The HIRES wavelengths have been divided
by the ratio of the wavelengths of HeII to HI Ly$\alpha$. 
The vertical ticks show the redshifts of detected 
C IV absorption systems. Flux is in
erg cm$^{-2}$ s$^{-1}$~\AA $^{-1}$.\label{fig:0302hiheciv}}
  \end{center}
\end{figure}

Another possibility advanced by Reimers et al. (1997) is that
\ion{He}{2} opacity gaps may be caused by discrete ionizing sources
along or near the QSO line of sight.  In this case, the UV source
would most likely be an AGN or QSO since star-forming galaxies will
not substantially ionize intergalactic \ion{He}{2} (\cite{FGS98}).
Support for a discrete ionizing source comes from the optical Keck
HIRES data of Q~0302--003.  As shown in Figure~\ref{fig:0302hiheciv},
the Keck spectrum reveals eight \ion{C}{4} absorption complexes in the
redshift range probed by STIS. The redshifts of these complexes are
indicated by vertical bars.  We follow Steidel (1990) in defining a
complex as a group of systems spanning less than 1000 \kms\, in
velocity space; in each of the relevant complexes towards Q~0302--003,
the components span less than 200 \kms.  Two of the complexes, at $z=$
3.22 and 3.27, are at the edge of (or within) the proximity zone of
the QSO. One \ion{C}{4} complex, at $z=3.0$, falls too close to the
\ion{H}{1} \Lya\, geo-coronal emission line to judge whether it has
associated transmission of \ion{He}{2} \Lya .  Of the remaining five,
four \ion{C}{4} absorption complexes, at $z=$ 2.79, 2.83, 2.97 and
3.05, are close to regions with diminished \ion{He}{2} \Lya\, opacity.
Assuming a Poissonian distribution of \ion{C}{4} systems, we estimate
a probability, $p = 12$\% of one coincidence of a \ion{C}{4} absorber
with a \ion{He}{2} opacity gap; four coincidences are much less likely
($p=0.02$\%).

\begin{figure}[htp]
  \begin{center}
    \plotfiddle{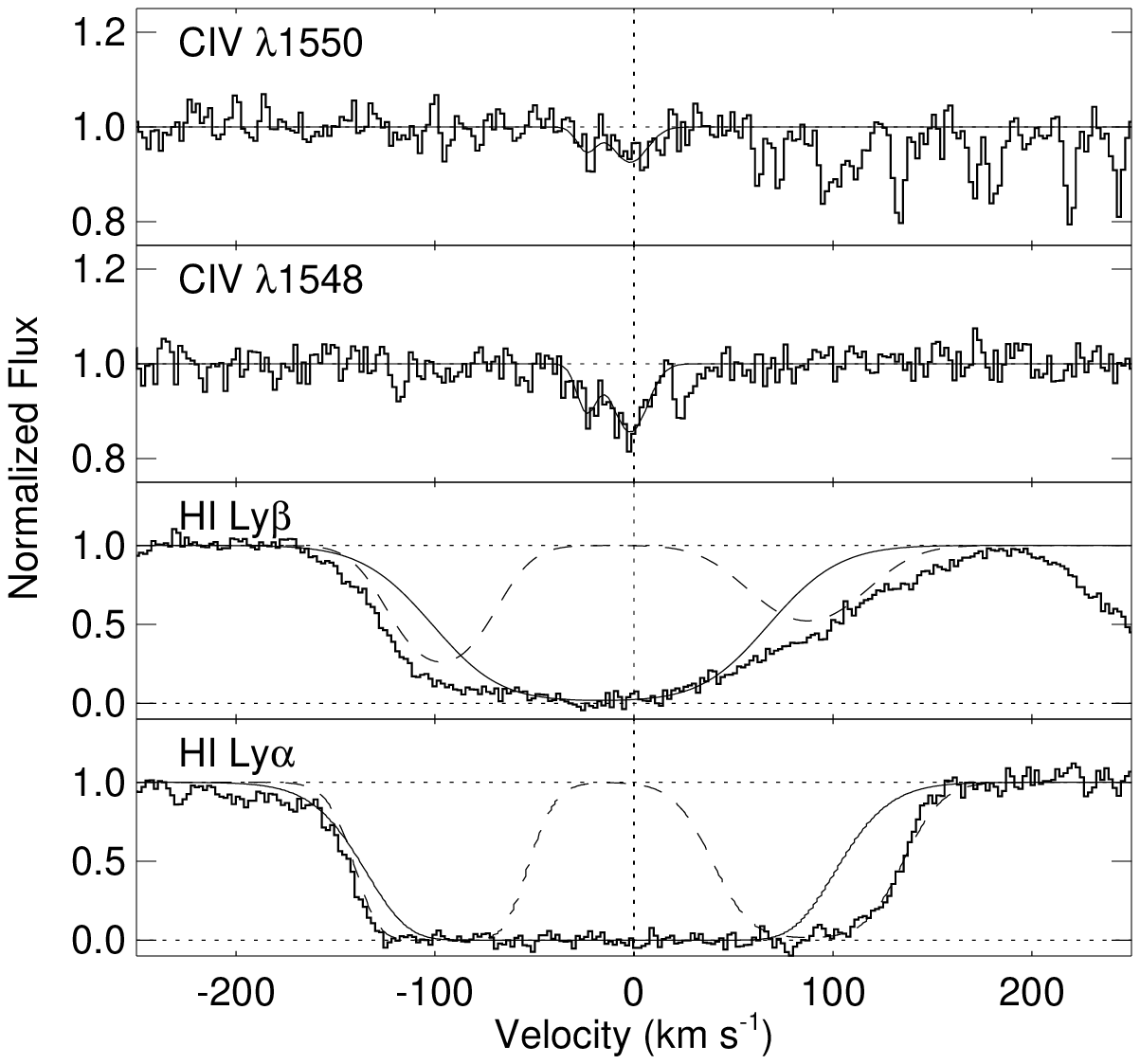}{16cm}{0}{135}{145}{-250}{-00}
    \caption{The Voigt profile 
fit in velocity space, in units of normalized flux vs. wavelength in \AA , to
the $z=3.047$ C IV doublet toward Q~0302--003. 
The \ion{H}{1} \Lya\,  and \Lyb\
regions required a three component fit, while the C IV doublet required only
two components.
\label{fig:0302z3047vpfit}}
  \end{center}
\end{figure}

The window at $z=3.05$ is of course the most prominent of the three.
Figure~\ref{fig:0302z3047vpfit} shows detailed plots of the \ion{C}{4}
and \ion{H}{1} absorption by this system.  Table~4 gives the detailed
measurements of absorbing systems near the major \ion{He}{2} opacity
gap at $z=3.05$. Successive columns list the ion (and spectral line),
measured redshift, Doppler parameter $b$, and column density.
The table also gives $4\sigma$ detection limits for a series of C, Si,
N and O ions.

Below, we test the hypothesis that the \ion{C}{4} absorber at $z=3.05$
could be photo-ionized by a hypothetical AGN that is also responsible for the
adjacent \ion{He}{2} opacity gap.  Our diagnostics include: the
required luminosity of the ionizing source, its spectral energy
distribution and the space density of galaxies or AGN at $z \sim 3$.  We
shall find consistent support for the discrete ionizing source
hypothesis.

\textit{Luminosity requirements}.  To first approximation, the
luminosity of the putative He$^+$-ionizing source relative to the QSO
scales with the relative volume of their Str\"omgren spheres. The
radius of the QSO Str\"omgren sphere is given by the physical extent
(luminosity distance in the local frame) of the proximity effect,
$\Delta \ell_{\mathrm prox} \approx 14.4$ Mpc, while the radius of the
ionized region caused by the putative ionizing source is the
half-width of the opacity gap $\Delta \ell_{\mathrm gap}\approx 2.15$
Mpc. Given these size estimates, the hypothetical source would need to
be only 0.3\% as luminous as the QSO ($V=17.4$, \cite{Veron98}) and
would have an apparent magnitude of $V=23.4$ if it were on the
line-of-sight to the QSO and radiates isotropically. A more detailed
luminosity estimate with Kraemer's (1985, \cite{Kraemer94})
photo-ionization code supports the conclusion that a bright Seyfert~I
galaxy about 0.3\% as luminous as Q~0302--003 could produce the
observed gap at $z=3.05$ if it were directly along the line of sight
to the QSO. A source offset from the line of sight would need to be
more luminous.

\textit{The ionizing flux distribution near the \ion{C}{4} absorber}.
Let us assume that the metal-line system at $z = 3.047$ originates in
the halo of the host galaxy (or a nearby galaxy in the host cluster)
of an AGN, the putative source of the ionization which produces the
opacity gap. To explore whether this \ion{C}{4} absorber could be
photo-ionized by the hypothetical AGN, we constructed photo-ionization
models with CLOUDY (\cite{Ferland98}).  For the model calculation, we
assumed the usual plane-parallel geometry and allowed the gas to be
ionized by a model AGN spectral energy distribution (SED) and/or a
Fardal et al. (1998) UV background at $z= 3.0$. The Fardal et al. UV
background spectra are more consistent with the results derived in \S
\ref{sec:modeling} than the Haardt \& Madau (1996) or Madau, Haardt \&
Rees (1999) ones because of their larger values for $S_{\mathrm L} =
J_{\mathrm H I}/J_{\mathrm He II}$. We tried the Fardal et al.
backgrounds due to QSOs with and without a contribution from stars
(see their Figure 6). We used Ferland et al.'s (1996) model of an AGN
SED, where the ``big blue bump'' is approximated as a power law with a
UV exponential cutoff with characteristic temperature $T_{\mathrm
  BB}$. We calculated the \ion{C}{4}/\ion{Si}{3},
\ion{C}{4}/\ion{Si}{4}, \ion{C}{4}/\ion{N}{5}, and
\ion{C}{4}/\ion{O}{6} column density ratios as a function of the
ionization parameter, $U$ = $n_{\gamma}/n_{\mathrm H}$, where
$n_{\gamma}$ is the \ion{H}{1} ionizing photon density and $n_{\mathrm
  H}$ is the total hydrogen number density.

\begin{figure}[htbp]
  \begin{center}
    \plotfiddle{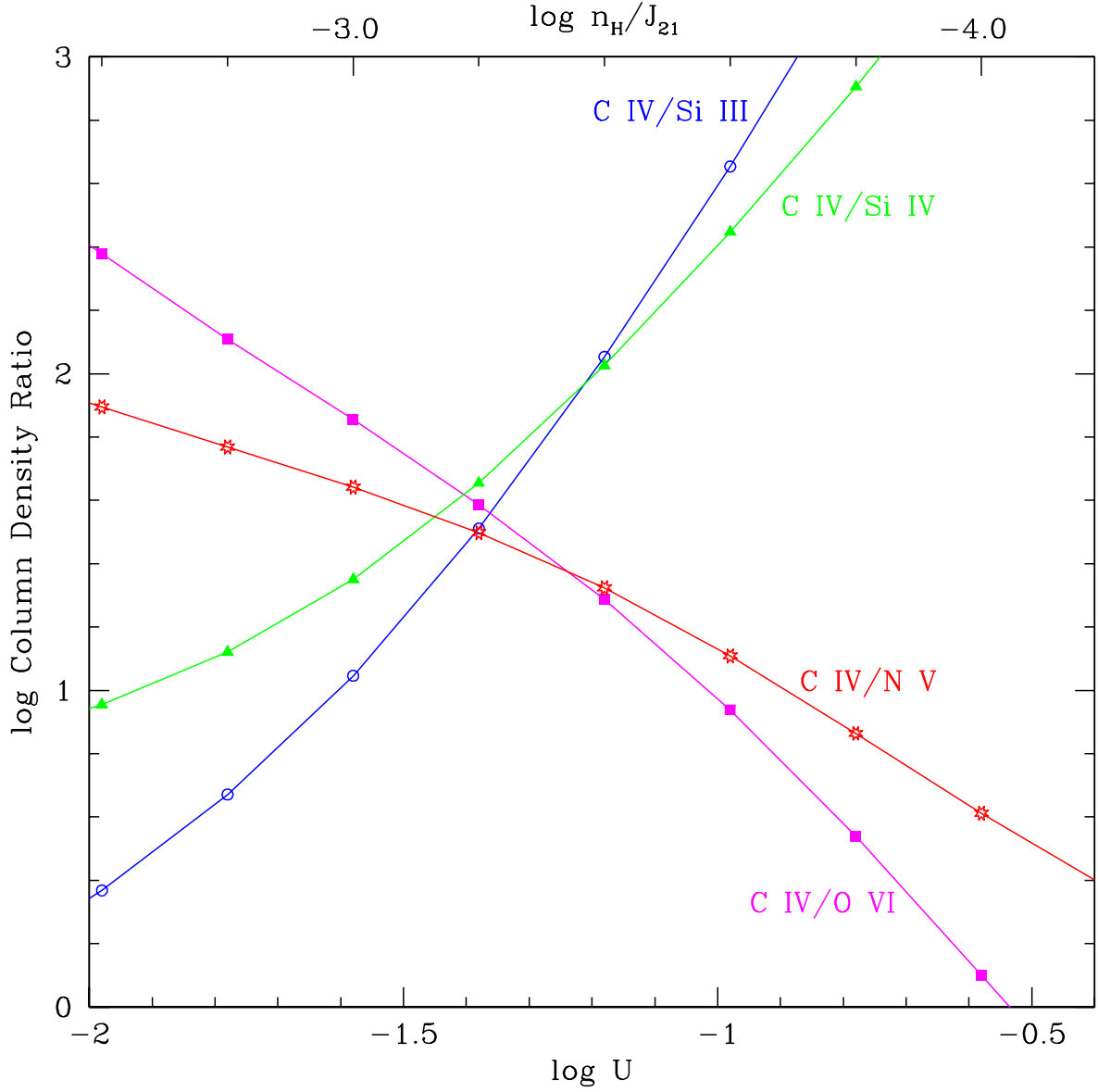}{15cm}{0}{85}{85}{-260}{-140}
    \caption{Column
  densities of high-ionization species, predicted by a
  photo-ionized gas model for the C IV absorber at $z_{\mathrm abs}$ =
  3.047 as a function of the ionization parameter $U$ (bottom axis)
  and total hydrogen number density $n_{\mathrm H}$ (top axis). This
  model assumes that the ionization is predominantly due to the UV
  background from QSOs and AGN and adopts the radiation field
  calculated by Fardal et al. (1998) for $z$ = 3.0. The relative heavy
  element abundances are assumed to be solar, and the overall
  metallicity [M/H] = -2.5 with log $N_{\mathrm HI}$ = 15.32.
\label{fig:tmt_fig}}
  \end{center}
\end{figure}

Figure~\ref{fig:tmt_fig} shows a sample CLOUDY calculation for the
case in which the ionizing flux is dominated by the UV background,
including a contribution from stars. We find that the model is in
agreement with the observed constraints (Table 4) for $10^{-1.30} \ 
\leq \ U \ \leq \ 10^{-0.76}$, which corresponds to 1.5$\times
10^{-4}$ $\leq \ n_{\mathrm H}/J_{21} \ \leq$ 5.3$\times 10^{-4}$
cm$^{-3}$ where $J_{21} = J_{\mathrm H I}/10^{-21}$ erg s$^{-1}$
cm$^{-2}$ Hz$^{-1}$ sr$^{-1}$. Since these are reasonable particle
densities for the distant outer halo of a galaxy, we conclude that it
is not {\it necessary} to have an AGN nearby to explain the observed
high level of ionization. However, we find from another CLOUDY
calculation in which the ionizing flux is due to a nearby AGN that the
model column density ratios can be brought into agreement with the
observations over a reasonable range of $U$ as long as the
characteristic temperature of the ``big bump'' $T_{\mathrm BB}$ is
less than $\sim 10^{5.5}$ K. Since luminous QSOs tend to have higher
values of $T_{\mathrm BB}$ (Hamann 1997), this indicates that the
putative AGN must be at least a low-luminosity QSO. We conclude that
it is possible, but not necessary, that a low-luminosity AGN near the
sight line at $z \approx$ 3.05 ionizes the region around the
\ion{He}{2} opacity gap and may also play an important role in the
ionization of the nearby \ion{C}{4} system.

\textit{The space density of opacity gaps vs. galaxies.}  How
plausible is it that regions of low \ion{He}{2} opacity are produced
by galaxies, or more likely, galaxies associated with AGN?  To answer
this question, we compare the number density of \ion{He}{2} opacity
gaps, $n_{\mathrm gap} \sim 5\times 10^{-5}$ Mpc$^{-3}$, with the
number density of star-forming galaxies and AGN as revealed in
narrow-band or emission-line surveys of galaxies at high-redshift.
These surveys, which are summarized in Table~5, show that the comoving
space density of galaxies at $2.4<z<3.4$ is $n_{\mathrm gal} \sim
60-410\times 10^{-5}$ Mpc$^{-3}$.  Hence, the space density of opacity
gaps is about 1--8 \% that of galaxies.  If the AGN fraction among
galaxies were similar, it would be consistent with the hypothesis that
\ion{He}{2} gaps are produced by such discrete ionizing sources.  The
incidence of low-luminosity AGN at high redshift is poorly known.
About 10\% of Lyman limit dropout galaxies at $z\sim 3$ are QSOs or
AGN (\cite{Steidel96}) with a space density of $\sim 1\times 10^{-4}$
comoving Mpc$^{-3}$.  An independent estimate by Teplitz et al. (1998)
from an extrapolation of the QSO luminosity function gives a similar
number density at $2.3<z<2.5$ for an H$\alpha$ emission line flux of
$10^{-16}$ erg \cms\,  s$^{-1}$.  This value is within a factor of two
of the number density of \ion{He}{2} low-opacity regions, so it is
quite plausible that the gaps are produced by AGN.


\subsection{Can the gaps be caused by shock-heated gas?}
\label{sec:shockheatedgas}

In hydro-dynamical cosmological simulations, galaxies are found in the
vicinity of hot, collisionally ionized gas regions.  This is not only
because the collapse of initial density perturbations leads to shock
heating of the gas; it also enables the condensation of cool objects
to form galaxies (Dav\'{e} et al. 1999; Cen \& Ostriker 1999). It is
likely that the \ion{C}{4} absorber near the \ion{He}{2} opacity gap
at $z=3.05$ arises in a galaxy halo.  Thus, it is possible that the
low \ion{He}{2} opacity is basically due to the high temperature of
the region in which the galaxy is located. The volume fraction of gas
with temperature 10$^{5} \  \leq \  T \  \leq 10^{7}$ K is roughly 0.01
at $z \approx $ 3 (see Figure 2 in Cen \& Ostriker 1999), which is a
factor of 4--10 times lower than that of the observed opacity gaps.
It may be possible to explain the discrepancy by uncertainties in the
hot gas fraction as computed by the hydro-dynamical codes; the matter
clearly deserves further study.

\subsection{Summary}

In summary, we have considered three explanations of the
high-transmission regions in the \ion{He}{2} Gunn-Peterson absorption
trough: (a) they result from underdense regions in a clumpy IGM; 
(b)  they are regions ionized by
nearby AGN; and (c) they arise in hot, collisionally ionized regions.  
We find that the observed opacity gap at $z=3.05$ is located in a region
unlikely to show an opacity gap caused by fluctuations in the diffuse IGM.
Discrete ionizing 
sources, however, are consistent
with the observed gaps for the following reasons:
(1) The luminosity required is
consistent with an AGN. (2) There is a \ion{C}{4} absorber at a
velocity separation of $\Delta v = 350$ \kms .  (3) The ionization
state of the \ion{C}{4} absorber does not preclude the existence of a
nearby AGN, although it does rule out the possibility of a bright QSO
in the immediate vicinity.  (4) The space density of \ion{He}{2} opacity gaps
is consistent with the best available estimates of AGN at $z\sim 3$.
Collisionally ionized regions, which are predicted in galaxy formation
models, do not produce a high enough filling factor of ionized gas in
current models.

\section{DISCUSSION AND CONCLUSIONS}
\label{sec:discussion}

Our analysis of the STIS spectrum of Q~0302--003 leads to the following
conclusions.

\noindent
\textit{1. Opacity of \ion{He}{2} \Lya\, at $z\sim 3$.}  Away from
distinct opacity gaps that we could identify in the He~II absorption,
we made direct measurements of the average transmission of the IGM.
We found values that ranged from $\bar{I}=0.0086$ for an interval
centered at $z=3.15$ to $\bar{I}=0.15$ for another zone centered at
$z=2.82$.  If the opacity were constant within one of these intervals
(which it is not), these intensities correspond to $\tau=4.75$ and
1.88, respectively.  A stronger statement can be made about the IGM
opacity when we note that there are no regions near $z=3.15$ that have
a relative transmission greater than 0.03.  If we apply this result to
the expectations for opacity fluctuations computed by Fardal et al.
(1998), we find that the average transmission must be even lower than
what we could measure, with representative values of optical depth
greater than 6.0, depending on some initial assumptions.

\noindent
\textit{2. Evolution of \ion{He}{2} $\tau$(\Lya ). }
The total \ion{He}{2} opacity rises more rapidly with
redshift than previously thought (e.g. \cite{FGS98}, Zhang et al. 1998).
The observed data are compatible with an opacity break occurring between
$z=2.9$ and $z=3.0$ as suggested by other lines of evidence 
presented by Songaila (1998). This rapid rise in
opacity, however, is overshadowed by the fact that at $z\sim 3$, 
\ion{He}{2} is such
a minor fraction of helium ($\sim 1.6\times 10^{-3}$). Clearly, 
reionization of helium is essentially complete by $z\sim 3$.

\noindent
\textit{3. Comparison with cosmological models.} Observations of  
\ion{He}{2} \Lya\,  absorption give  a direct connection with
cosmological simulations if differences in resolution are taken into 
account. A comparison of Zhang et al.'s model to the observations
indicates that their predicted \ion{He}{2} \Lya\ opacity is 3--5 times
lower than observed by STIS.
 
\noindent
\textit{4. The UV background and the \ion{He}{2} \Lya\, opacity.}  We
constructed detailed models of the Gunn-Peterson trough over the
observed range in redshift.  In agreement with Hogan et al. (1997), we
find that the observed proximity effect cannot be reproduced in the
presence of a diffuse component. Instead, a model based only on known
\ion{H}{1} \Lya\, lines provides a good match to the observations. The
model also requires that the UV background has a soft spectrum with a
softness parameter $S = \Gamma_{\mathrm HI}^{\mathrm J}/
\Gamma_{\mathrm HeII}^{\mathrm J}$ certainly larger than 400 and close
to 800, indicating a significant stellar contribution.  We can also
set the constraint that if $J_{\mathrm HI} \simlt 10^{-21}~ {\mathrm
  erg}~{\mathrm s}^{-1}~{\mathrm cm}^{-2}~{\mathrm Hz}^{-1}~{\mathrm
  sr}^{-1}$, then $S > 4000$.

Outside of the proximity region, most of the \ion{He}{2} Gunn-Peterson
trough requires a large value of $\eta = N_{\mathrm HeII}/N_{\mathrm
  HI} \simeq 350\pm 50$, which in turn requires both a diffuse gas
component and a soft UV background with values of $S \sim 800$. In
particular, these constraints are needed to explain the high
\ion{He}{2} opacity in the Dobrzycki-Bechtold void.  As also deduced
from the observations of HS 1700+6416 (Davidsen et al.  1996), our
model indicates that the UV background has a harder spectrum at $z <
3$ as the Universe becomes transparent to \ion{He}{2} Lyman continuum
radiation.

\noindent
\textit{5. Nature of the opacity gaps.}  
We considered three explanations for the \heii\, opacity gaps observed
in QSO spectra: shock-heated gas, low density regions demarcated by
gaps in the \hi\, \Lya\, forest, and regions ionized by a discrete,
local sources.  The shock-heated gas scenario, as calculated by
current hydro-dynamical codes, is inconsistent with the observed
opacity gap filling factor.  The clumpy IGM model of Miralda-Escud\'e,
Haehnelt \& Rees (1999) predicts low-opacity regions with a mean free
path similar to that of the observed \heii\, opacity gaps.  
We can reproduce the
observed \heii\, opacity gaps only if $\eta\simlt 100$, contrary to
other regions of the spectrum where $\eta \simeq 350$.  The
softness parameter 
corresponding to $\eta\simlt 100$,
$S\la 200$, is
consistent with a local hard ionizing source, such as
an AGN.

In the case of the $z=3.05$ opacity gap, there are further supporting
reasons for the presence of such an ionizing source.  (1) The
luminosity required is consistent with a bright Seyfert I galaxy.  (2)
There is a nearby \ion{C}{4} absorber at a velocity separation of
$\Delta v = 350$ \kms , possibly indicating the existence of a nearby
AGN. Three
other regions of lower \ion{He}{2} opacity 
can also be associated with a \ion{C}{4} system. We estimate
at $p = 0.02\%$ the probability that four opacity gaps are associated
with \ion{C}{4} absorbers by chance.  (3) The ionization state of the
\ion{C}{4} absorber does not preclude the existence of a nearby AGN.
(4) The space density of \ion{He}{2} gaps ($5 \times 10^{-5}$ comoving
Mpc$^{-3}$) is consistent with the best available estimates of AGN at
$z\sim 3$.

\acknowledgements

We are most grateful to Toni Songaila and to Michael Rauch and Wal Sargent
for providing us with Keck HIRES spectra of Q~0302--003.
We would also like to acknowledge the following people for their help with this
study:  
E. Agol for providing the program KERRTRANS;
K. Feggans and C. Sehman for developing some IDL routines for CLOUDSPEC;
Steve Kraemer, for useful discussions and running some
   photo-ionization models; 
Don Lindler, for software; 
Harry Teplitz, for useful discussions; 
James Wadsley, for useful discussions and running some
hydro-dynamical models; 
Theodore Gull and Steve Maran who contributed to the manuscript.
This work was supported by NASA through funding of the STIS IDT.

\end{document}